\documentclass[twocolumn]{aastex62}

\usepackage{amsmath}	% Advanced maths commands

\newcommand{\RM}{radial migration}
\newcommand{\Rm}{Radial migration}
\newcommand{\ct}{cold torquing}
\newcommand{\Ct}{Cold torquing}
\newcommand{\CR}{corotation}

\newcommand{\ULR}{ultraharmonic}
\newcommand{\pesc}{PESC}
\newcommand{\Fe}{metallicity}
\newcommand{\RMSL}{$\langle(\Delta L_z)^2\rangle^{1/2}$}
\newcommand{\DelEran}{$|\Delta E_{\rm ran}|^{1/2}$}
\newcommand{\kmy}{kinematically}
\newcommand{\gc}{galactocentric}
\newcommand{\anu}{annulus}
\newcommand{\epi}{epicyclic}
\newcommand{\axisym}{axisymmetric}

\usepackage{xcolor}
\definecolor{dcolor}{RGB}{50,205,50}
\definecolor{kcolor}{RGB}{255,10,25}
\definecolor{sarahcolor}{RGB}{167, 66, 244}

\received{April 15, 2019}
\revised{??}
\accepted{July 18, 2019}
\submitjournal{ApJ}

\shorttitle{When Cold Migration is Hot}
\shortauthors{Daniel et al.}

\begin{document}

\title{When Cold Radial Migration is Hot: Constraints from Resonant Overlap}

\correspondingauthor{Kathryne J. Daniel}
\email{kjdaniel@brynmawr.edu}
\author[0000-0003-2594-8052]{Kathryne J. Daniel}
\affil{Bryn Mawr College, Department of Physics, 101 N Merion Ave., Bryn Mawr, PA 19010, USA}

\author[0000-0002-9180-6565]{David A. Schaffner}
\affil{Bryn Mawr College, Department of Physics, 101 N Merion Ave., Bryn Mawr, PA 19010, USA}

\author{Fiona McCluskey}
\affil{Bryn Mawr College, Department of Physics, 101 N Merion Ave., Bryn Mawr, PA 19010, USA}

\author{Codie Fiedler Kawaguchi}
\affiliation{Los Alamos National Labs, PO Box 1663 MS E526, Los Alamos, NM 87545, USA}

\author[0000-0003-3217-5967]{Sarah Loebman}
\altaffiliation{Hubble fellow}
\affiliation{Department of Physics, University of California, Davis,1 Shields Ave, Davis, CA 95616, USA}

%%%%%%%%%%%%%%%%%%%%%%%%%% ABSTRACT %%%%%%%%%%%%%%%%%%%%%%%%%%%
\begin{abstract}
It is widely accepted that stars in a spiral disk, like the Milky Way's, can radially migrate on order a scale length over the disk's lifetime. With the exception of cold torquing, also known as \lq\lq churning," processes that contribute to the radial migration of stars are necessarily associated with kinematic heating. Additionally, it is an open question whether or not an episode of cold torquing is kinemically cold over long radial distances. This study uses a suite of analytically based simulations to investigate the dynamical response when stars are subject to cold torquing and are also resonant with an ultraharmonic.  Model results demonstrate that these populations are kinematically heated and have RMS changes in orbital angular momentum around corotation that can exceed those of populations that do not experience resonant overlap.  Thus, kinematic heating can occur during episodes of cold torquing. In a case study of a Milky Way-like disk with an exponential surface density profile and flat rotation curve, up to 40\% of cold torqued stars in the solar cylinder experience resonant overlap. This fraction increases toward the galactic center. To first approximation, the maximum radial excursions from cold torquing depend only on the strength of the spiral pattern and the underlying rotation curve.   This work places an upper limit to these excursions to be the distance between the ultraharmonics, otherwise radial migration near corotation can kinematically heat. The diffusion rate for kinematically cold radial migration is thus constrained by limiting the step size in the random walk approximation.
\end{abstract}

%%%%%%%%%%%%%%%%%%%%%%%%%% KEYWORDS %%%%%%%%%%%%%%%%%%%%%%%%%% 
% A maximum of six subject keywords should be listed, in alphabetical order, after the abstract.

%%%%%%%%%%%%%%%%%%%%%%%%%% TEXT %%%%%%%%%%%%%%%%%%%%%%%%%% 
\section{Introduction} \label{sec:intro}

Transient spiral arms drive a range of dynamical processes of significant importance to the internal, secular evolution of galactic stellar disks \citep[see][and references therein for a review]{Sellwood14}.  
Many of these processes cause disk stars to shift radial position away from their birth radius over time, as their orbital angular 
momenta,\footnote{A change in a star's orbital angular momentum is associated with a change in its mean orbital radius
as long as the radial circular velocity profile of the disk, which is governed by the potential, is not $\propto R^{-1}$.  Otherwise, any change in orbital angular momentum will necessarily result in a change in the size of the orbit.} 
orbital energies, and eccentricities can be considerably altered.  
Such changes are often broadly attributed to ``\RM,'' a term which does not distinguish between the manner of change to radial position.  
In fact, most physical processes that induce \RM\,necessitate simultaneous changes to both orbital angular momentum and orbital circularity \citep[e.g.,][]{SS53,BW67,LBK72,CS85}.  
However, one particular process can reorganize orbital angular momentum in the disk and is assumed to never be associated with kinematic heating, hereafter called `` \ct .''\footnote{The terminology adopted here is the result of a discussion held at the Aspen Center for Physics in August 2018 to determine a standard nomenclature for various mechanism that radially mix stellar populations.}
Cold torquing is the physical mechanism identified by \cite{SB02} that drives \lq churning' of stellar populations.
A star can migrate by \ct\,when it is in a stable orbit with the \CR\ resonance of a transient spiral; the \CR\ resonance occurs where the circular orbital frequency of stars equals the pattern speed of a spiral (or other) perturbation to the potential.

Present day kinematic, chemical, and structural properties of the stellar disk depend on the amount of kinematic heating produced by all forms of \RM\ over time.
The degree to which \kmy\,cold forms of \RM\,affect the evolution of disk galaxies is thought to be directly dependent on the efficiency of \ct.  
However, theoretical studies have yet to develop a deep understanding for how \kmy\ cold \RM\ through \ct\ truly is in all circumstances.
It is therefore important to reevaluate the assumption that \ct\,is always \kmy\,cold and, if not, to understand any constraints to that assumption.

In order for \ct\ to be efficient, multiple generations of transient spiral arms with multiple pattern speeds (and thus \CR\,radii) must occur over a disk's lifetime.  
Each episode causes stars to take a single step in a random walk-like redistribution of orbital angular momentum, where the standard deviation in the final distribution of stars for a given birth radius is proportional to the size of a single step.  Thus, a limit to the size of a single step in the radial redistribution of stars via \ct\ sets a limit to the final redistribution of stars over the lifetime of the disk. 

It is generally accepted that the maximum radial distance a star can migrate via a single episode of \ct\ \citep{SB02}, 
\begin{equation}\label{eqn:Rmax}
	\Delta R_{\rm max} = 2 \sqrt{\dfrac{|\Phi_1|}{AB}},
\end{equation}
depends on the square root of the amplitude of the perturbation to the potential, $\Phi_1$, and the Oort constants for sheer and vorticity, $A$ and $B$.
The above expression reduces to $\Delta R_{\rm max} \propto \sqrt{|\Phi_1|}$ in a disk with a flat rotation curve.
Under this assumption, stars can have arbitrarily large changes in their orbital angular momenta with no change to their orbital eccentricity since the amplitude of these changes \textit{only} depends on the spiral strength.  

It has long been accepted that additional patterns exist in the disk, like a bar, and these perturbations each have resonances, like the Lindblad resonances (LRs) and their \ULR s.  \cite{MF10,Minchev11} recognized that the radial range for \RM\ could be enhanced by overlapping a harmonic of a LR from one pattern with the \CR\ resonance of another.  They found that stellar populations that were subject to this type of resonant overlap had strong signatures of \RM\ since the RMS changes in orbital angular momentum in regions of resonant overlap were greater than the expected sum from each resonance, thus suggesting a non-linear, enhanced disk response.   

It is usually presumed that a single spiral pattern could not have regions of resonant overlap in the phase-space of a stellar population. However, \cite{DW15} showed that the \CR\,resonance is not defined by a single radius. Rather, \textit{\CR\ is better described by a region} in phase-space which can overlap with \ULR s from the same pattern.  \cite{SD19} used a novel approach, called Permutation Entropy and Statistical Complexity (\pesc)~\citep{Bandt2002,Rosso2007,Weck2015,brown2015,Schaffner2016}, to identify the dynamical response for stars meeting both the \CR\,and \ULR\,resonant criteria \textit{from a single perturbation} to be chaotic.  

This paper investigates how the overlap of the \CR\ region with the inner and outer \ULR\ resonances of the same perturbation affects disk kinematics.     
A review of the theory relevant to dynamical resonance in a disk, \ct, and resonant overlap is given in \S\ref{s:theory}. The approach is described in \S\ref{s:models}, including the model used (\S\ref{s:ICs}), the production of orbits (\S\ref{s:simulation}), the quantitative characterization of those orbits (\S\ref{s:trapping}), and the orbital categorization scheme used for analysis (\S\ref{s:classification}).  
This study is limited to the disk response from a single transient spiral pattern, but a discussion on scaling relations for the degree of kinematic heating associated with \ct\ is given in \S\ref{sec:discussion}.  
The discussion in \S\ref{s:heating} proposes a reevaluation of the assumption that \ct\,from a single spiral pattern is necessarily a cold process.  
The discussion \S\ref{s:DeltaRmax}, considers limits to the radial range within which \ct\,is \kmy\,cold.
A brief summary of the conclusions are given in \S\ref{s:conclusion}.

\section{Theoretical Background}\label{s:theory}

The significance of \ct, as opposed to other mechanisms for \RM, is that it is assumed to be \kmy\,cold.  This result arises from the derivation for \ct\,in action space since it conserves the radial action of a star while altering azimuthal action (equal to angular momentum in a disk) \citep{SB02}.  For a population of thin disk stars, with orbital eccentricities small enough that the \epi\ approximation holds, this translates to altering orbital sizes without increasing the radial velocity dispersion.  Populations of stars that are subject to \ct\ also have, on average, conserved vertical action \citep{SSS12} as long as non-\axisym\,perturbations are relatively weak and the affected orbits only mildly eccentric \citep{VCdO16}. 

The impact of \ct\,on disk evolution is different from the impact from forms of \RM\,that induce kinematic heating, but, in practice, it is difficult to disentangle the relative importance of each.
Large surveys of Milky Way stars have produced significant evidence for \RM\,including highly suggestive tracers for past episodes of \ct.  \Rm\ is frequently invoked to explain the degree of increasing spread in the \Fe\ distribution at a given \gc\,radius with increasing age \citep[e.g.,][]{Casagrande11,Ness16} but the extent to which this can be attributed to \ct\ is unclear.  The [Fe/H] distributions from APOGEE \citep{Alam15MNRAS} for stars in the plane of the disk (vertical height $|z|<0.50$~kpc) are skewed with tails toward lower/higher \Fe\ in the inner/outer disk \citep{Hayden15}.  Such a skew can be understood to arise when stars from the outer/inner disk have significant changes in their orbital angular momentum and thus, assuming a more rapid star formation rate in the inner disk, contaminate stellar populations in the inner/outer disk with lower/higher \Fe\ stars.   Indeed, these skews were fit by a N-body+SPH simulation that resolves \RM\ including \ct\ \citep{Loebman16}.  However, a recent model that treats all types of \RM\ with a single diffusive prescription is also able to reproduce a similar skew \citep{Frankel18}.   
Adding kinematic information can greatly help constrain the role of \ct.  Perhaps the smoking gun for past \ct\ in the Galaxy comes from the RAVE survey \citep{Steinmetz06} with which \cite{Kordopatis15} identified metal enhanced stars in the solar neighborhood on nearly circular orbits. In fact, the \Fe\,of the Sun suggests it has migrated from a birth radius that was on order a scale length closer to the galactic center \citep{WFD96,Frankel18}.

It has been proposed that very efficient \ct\ over the lifetime of the disk could even lead to the emergence of structures like the outer disk \citep{Roskar08a,DRL17} and the thick disk \citep{SB09b,Loebman11,SM17}.  
Attempts to theoretically constrain the past efficiency of \ct\ in the solar neighborhood have had moderate success.
A model that included a prescription for \ct\ \citep{SB09b,SB09a} was able to reproduce solar neighborhood chemistry in the thin disk.  However, this model treated \ct\,as a diffusive process that did not depend on the kinematic temperature of the affected population. 
\cite{DW18} used an analytic approach to demonstrate that the fraction of stars that could migrate radially via \ct\,decreases with increasing velocity dispersion.  This is in agreement with results from simulations which find preferential \ct\ for populations with smaller velocity dispersion and vertical excursions \citep{SSS12,VCdON16}.  
N-body simulations of a quiescent MW-like disk \citep{ABS17} found that the redistribution of angular momentum for stars born in a region corresponding to the Solar neighborhood nearly matched \citeauthor{SB09b}'s prediction.  However, while many models and simulations are able to reproduce Solar neighborhood chemistry and kinematics, the explanations are sometimes seemingly contradictory.  This likely points to a lack of complete understanding for the dynamical mechanisms driving \RM.

Analytic scaling relations that can use spiral disk structure to place limits on the impact of \ct\,could assist our interpretation of observational and simulated data.

\subsection{Corotation Resonance}

A particular family of orbits, sometimes called \lq horseshoe' orbits, underlie the physics that can lead to \ct\ \citep{GT82,SB02,DW15}.  These orbits occur near the \CR\ resonance, where the orbital frequency of stars, $\Omega(R)$, equals the pattern speed, $\Omega_p$, of a non-axisymmetry in the disk.  A star with a trajectory that belongs to this orbital family, hereafter called \lq trapped', will have periodic changes in orbital angular momentum, $L_z$, causing its mean orbital radius, $R_L$, to oscillate about the radius of \CR, $R_{\rm CR}$.  The mean orbital radius of a star can be defined by using its orbital angular momentum such that,
\begin{equation}\label{eqn:RL}
	R_L = \dfrac{v_\phi}{v_c} R,  
\end{equation}
where $v_c$ is the orbital circular velocity at $R_L$, $v_\phi$ is the instantaneous tangential velocity about the disk, and $R$ is the instantaneous radial position of the star. Critical to the definition for \ct\,, there is little to no change in the star's orbital circularity after a trapped orbital period.  Should the non-\axisym\ pattern be transient, a trapped star could have a permanent change in its orbital angular momentum, and thus mean orbital radius, and no change in its orbital circularity \citep{SB02}.  
  
An approximation for the behaviour of a star in a trapped orbit was derived in Section~3.3.3 of \cite{BT87} and invoked to describe limits on \ct\ by \cite{SB02}.  This approximation assumes a smooth disk that is perturbed by a weak bar pattern with potential amplitude, $|\Phi_b|$, at the radius of \CR.  By making the further assumption that the underlying disk has a flat rotation curve, the \textit{maximum} radial excursion for a star in a trapped orbit scales as $\Delta R \propto \sqrt{|\Phi_b|}$, and the \textit{minimum} time-scale for the \textit{smallest} excursions scale as $T_{min} \propto R_{\rm CR}/\sqrt{|\Phi_b|}$.  While these provide guidelines for approximating the efficiency of \ct, it may be informative to refine these scaling relations.
 
\cite{DW15} demonstrated that the location of \CR\ does not occur only at the radius where $\Omega(R)=\Omega_p$. Rather \CR\ can be better described in coordinate space by a 2D region in the plane of the disk.  Disk stars with their mean orbital radius within this \lq \CR\ 
region' are, to first order, trapped in stable orbits around \CR.  
The analytic criterion is derived in action space where vertical action is assumed to be separable in a thin disk.  A transformation can be made from 4D action-angle-space to 4D phase-space for a given radially local rotation curve, spiral strength, and pitch angle.  A star is trapped when it's mean orbital radius, $R_L$, is approximately within the 2D coordinate-space \CR\,region, where there is a higher order dependence on orbital circularity.
The shape of the \CR\ region depends on the morphology of the spiral pattern in that the radial range of the \CR\ region increases with increasing spiral strength and openness of the spiral arms.  There is also a dependence on the rate of divergence with radial distance from \CR\ between the spiral arm's pattern speed and the orbital frequency for disk stars.  For example, spirals that corotate with the disk at all radii ($\Omega(R)=\Omega_p(R)$) have the largest \CR\ regions, possibly spanning the full radius of the disk \citep{GKC12,DVH13}, while a spiral with radially constant pattern speed ($\Omega_p=const$) in a Keplerian disk ($\Omega(R) \propto R^{-3/2}$) will have a much less extended \CR\ region \citep[see equation 32 in][for their generalized analytic expression for the shape of the \CR\ region]{DW15}.

\subsection{Lindblad Resonances and the Ultraharmonics}

The inner and outer Lindblad resonances (ILR/OLR) are where a disk star passes or is passed by the spiral pattern at the star's \epi\ frequency, $\kappa$.  The LRs and their harmonics occur at radii where,
\begin{equation}
	\kappa = \pm m\,(n+1) \,[\Omega_p-\Omega(R)],
\end{equation}
is satisfied, where $n$ stands for the $n^{th}$ harmonic of the Lindblad resonance and $m$ is the number of spiral arms.  The inner and outer Lindblad resonances have harmonic number $n=0$ in this notation.
In a disk with a flat rotation curve, the radial locations for the LRs are given by,
\begin{equation}\label{eqn:LR}
	R_{\rm LR}^{(n+1)} = \left(1 \mp \dfrac{\sqrt{2}}{m\,(n+1)} \right) \dfrac{v_c}{\Omega_{p}},
\end{equation}
where $\Omega_{p}=v_c/R_{\rm CR}$.

\subsection{Resonant Overlap}

Higher order harmonics of the LRs ($n>0$) are alone not expected to have a significant impact on disk kinematics.  However, since the \CR\ region has some finite radial range, it is possible for stars in otherwise stable, trapped orbits that have radially periodic orbits about the \CR\ radius to temporarily also be in resonance with a harmonic of the LRs.  
\cite{Chirikov79} predicted that 
stochastic\footnote{For clarity, in this paper, we reserve the use of the term {\it stochastic} to mean noisy or random, particularly in the quantum mechanical sense, and use {\it chaotic} or {\it complex} when referring to seemingly random behaviour that arises from complicated non-linear but deterministic processes. It is likely that use of the term \lq stochastic' by \cite{Chirikov79} refers more to the chaotic connotation than true stochasticity.} 
behaviour emerges in cases where more than one resonance occupies the same N-dimensional space.  He drew particular attention to the case of a pendulum under the influence of an external, periodic perturbation; a well known example of chaotic behaviour.  Explorations into the kinematic response of a stellar disk in the presence of multiple, non-\axisym\ patterns with different pattern speeds suggest there is a stochastic response for stars at radii that meet a resonant criterion for each pattern \citep[e.g.,][]{Quillen03,Jalali08,MF10}. 

The nature and degree of the dynamical response to resonant overlap can be challenging to identify. \lq Wild' \citep{Martinet74} or ergotic \citep{Athanassoula83} behaviour associated with resonant overlap can be identified in regions of a surface of section (SoS) diagram by irregularly distributed consequents \citep[see][\S 3.7.3]{BT08}.  \cite{Pichardo03} combined SoS and Lyapunov analyses to show that for a sufficiently strong perturbation, orbits in these regions exhibit chaotic behaviour.  SoS analysis can have limited utility since, in order to  to resolve a chaotic signature, it necessitates a significant number of orbits in order to populate the phase-space within a given energy slice and these are best evolved over several dynamical periods.

\begin{figure}
	\includegraphics[width=\columnwidth]{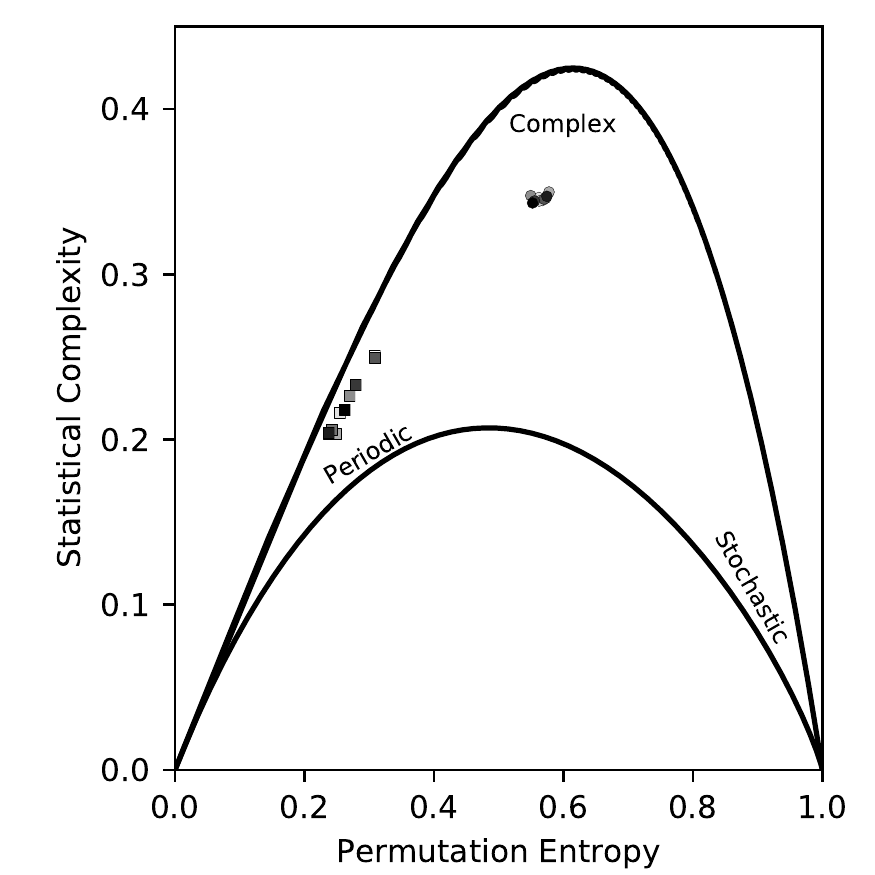}\\
    \caption{Illustrative example of the bi-modal distribution apparent from \pesc\,analysis for M6e from \cite{SD19} for orbits trapped at \CR\ that do or don't have an episode of resonant overlap.  Orbits are analyzed over 200~Myr ($0.85 \,T_{\rm Dyn}$) using snapshots spaced by 9~Myr ($T_{\rm Dyn}/19$).  Orbital analysis is done in groupings of 300 orbits.  Shading, from lighter to darker, indicates lower to higher initial orbital angular momentum, respectively.  There is no clear trend in the dynamical response within a given orbital type as a function of initial $L_z$. Orbits experiencing resonant overlap (circles, Type~3$\rightarrow$2) clearly occupy a region of the \pesc\,diagram that is indicative of complexity or classical chaos, while orbits trapped at \CR\,only (squares, Type~1) occupy a region that describes periodic orbits. {A general overview of the technique and its applications is presented in \cite{Weck2015} and \cite{brown2015}, and a description of the delay variation technique used on a different timeseries dataset is discussed in \cite{Schaffner2016}}.  Complete definitions for the parameters used and methods {for the particular example of galactic orbits} are deferred to \cite{SD19}.}
    \label{fig:CH}
\end{figure}

In a simultaneous study to the present paper, \cite{SD19} use \pesc\ analysis to effectively investigate the nature of the dynamical response in cases where the \CR\,region overlaps an \ULR\ produced by the same perturbation.
Since perturbation theory cannot predict non-linear effects from resonant overlap, \pesc\ analysis provides an avenue to identify such a dynamical response.
The \pesc\ method is well suited to identifying chaos in poorly resolved or limited data-sets since it does not need to fill the phase-space, it best identifies a chaotic response on shorter, rather than longer, time-scales,
and it uses only one governing dimension.  
%Neither SoS nor \pesc\ analysis require knowledge of the underlying potential.  The point here is that although the orbits in this study are evolved through a known potential, this is not a requirement to use the method. \pesc\ analysis can be used without any understanding of the driver for a chaotic region in the space under investigation.
In the case of resonant overlap, time-scales from classical perturbation theory do not apply since non-linear affects causing the chaotic response happen on arbitrarily short time-scales.
For some of the same models described in this work, \citeauthor{SD19} find a significant chaotic signature using only orbital angular momentum for star particles in small batches of few$\times10^2$ orbits on time-scales less than an orbital period and time-resolution equal to $\sim T_{\rm Dyn}/20$. 
Figure~\ref{fig:CH} shows an example of the results using \pesc\ analysis for orbits that are resonant with only \CR\ (squares, Type~1) and orbits that are resonant with both \CR\ and an \ULR\ (circles, Type~3$\rightarrow$2) in model M6e (described below in \S\ref{s:models}).
There are two relevant points. First, there can be a chaotic dynamical response even for a single spiral pattern. Second, that chaotic response is readily quantifiable.

The nature of the dynamical response during resonant overlap is not the aim of the present study.  Rather, the focus is on whether or not the disk response due to resonant overlap can place an upper limit on the efficiency of \ct. 
In the case where the \CR\,region overlaps the $1^{st}$ harmonic of the 
LR any stochastic or chaotic dynamical response is of interest.  
The timescale for a response at the \ULR\ is reasonable since our tracer particle simulations show that a trapped orbital period is on order a few \epi\ periods.
Such a response minimally implies that trapped stars with excursions in their mean orbital radii ($|\Delta R_L|$) greater than the distance between the $n=1$ harmonic of the LRs,
\begin{equation}\label{eqn:DeltaULR}
	\Delta R^{(2)}_{\rm LR} = \dfrac{\sqrt{2} R_{\rm CR}}{m},
\end{equation}
would have a chaotic response.  A strong dynamical response at resonant overlap could lead to a change in orbital family from trapped orbits to non-trapped orbital families.  A chaotic response is also a likely driver for \RM\ that is associated with kinematic heating.  Since kinematic heating is not expected to be associated with \RM\ around \CR\ such resonant overlap is of significant interest. 

\section{Models}\label{s:models}

Several models are used to explore the dynamical response due to resonant overlap.  All models assume a $m=4$ armed spiral pattern in a 2D disk potential that produces a flat rotation curve.  Within these potentials a set of orbits is evolved.  The method for producing each model's set of orbital initial conditions is discussed in \S\ref{s:ICs}.  These initial conditions are evolved to produce a trajectory as described in \S\ref{s:simulation}.  Each trajectory is then categorized as described in \S\ref{s:trapping}-\ref{s:classification}.

\subsection{Initial Conditions}\label{s:ICs}

Initial conditions for each star particle is produced by sampling the distribution function, $f_{\rm new}$ from \citet{Dehnen99}.  
This distribution function resembles a \kmy\ warm 2D disk and has moments that are similar to observed trends in the Milky Way; namely, a flat rotation curve \citep[e.g.,][and references therein]{Rubin83,Sofue09} and an exponential surface density profile \citep[e.g.,][]{Freeman70,vanderKruit87,Juric08,deJong10}. 
In energy-momentum space this distribution function is given by,
\begin{equation}\label{eqn:fnew}
f_{\rm new}(E,L_z) = \dfrac{\Sigma(R_E)}{\sqrt{2}\pi \sigma_R^2(R_E)} \exp\left\lbrace\dfrac{\Omega(R_E)[L_z-L_c(R_E)]}{\sigma_R^2(R_E)}\right\rbrace,
\end{equation}
where $R_E$ is the orbital radius for a star in a circular orbit with energy $E$, $\Omega(R)$ is the circular frequency at a given radial coordinate, $L_z$ represents orbital angular momentum about the $z$-axis, $L_c(R)$ is the orbital angular momentum for a star in a circular orbit at radius $R$, and $\sigma_R(R)$ is the radial velocity dispersion at radius $R$.

The \texttt{Python} based galaxy modelling package \texttt{galpy v1.4} \citep{Bovy15}\footnote{The galpy package can be accessed using the following link: http://github.com/jobovy/galpy.} is used to produce a set of initial conditions for each model. The \texttt{galpy} command \texttt{dehnendf} is set to numerically refine the initial distribution function given by equation~\ref{eqn:fnew} over 20 iterations so that the final distribution function better approximates an exponential surface density profile and flat rotation curve \citep[see][for more details]{Bovy15}.  This is done by setting $\beta=0$ and \texttt{niter}$=20$.
\texttt{galpy} uses natural length unit $h_0$ and natural velocity unit is $v_0$.
Both the radial velocity dispersion profile ($\sigma_R(R)$) and the radial surface density profile, $\Sigma(R)$, are assumed to be exponential.  
The scale length, $h_\sigma$, for $\sigma_R(R)$ is set to $h_\sigma=h_0$ and is assumed to be three times the scale length, $h_R$, for $\Sigma(R)$.  The radial velocity dispersion profile is normalized so that $\sigma_R(h_0) = 0.16\, v_0$.  

The \texttt{galpy} function \texttt{sample} is used to produce initial conditions for a set of test particles in several models. 
The number of initial conditions, $N$, minimum and maximum sampling radii, $R_{min}$ and $R_{\rm max}$ respectively, for each model are given in Table~\ref{tbl:3models}.  The radial ranges for the sampling annuli are chosen to span approximately 4~kpc centered on \CR\ in order to provide a complete sampling of the \CR\ region with minimal inclusion of initial conditions for stars that are never trapped at the \CR\,resonance.

4D phase-space coordinates were recovered for each initial condition by setting parameter \texttt{returnOrbit=True}.  Each phase-space coordinate is transformed from natural to physical coordinates using $v_0=220$~km~s$^{-1}$ and $h_0=8$~kpc.

\subsection{Orbital Integration}\label{s:simulation}

Each orbital trajectory is calculated from the initial phase-space coordinate using the 2$^{nd}$ order leapfrog orbital integrator with fixed $\delta t = 10^5$~yr time-steps initially described in \citet{DW15}. Orbits are evolved through the potential given by,
\begin{equation}
	\Phi(R,\phi,t)=\Phi_0(R) + \Phi_1(R,\phi,t), 
\end{equation}
where $\Phi_0(R)$ is the radially dependent \axisym\ disk potential and $\Phi_1(R,\phi,t)$ is a time-dependent non-\axisym\ spiral perturbation to the underlying potential.  

An analytic form for the underlying potential is chosen such that it is approximately consistent with the moments produced by the adopted distribution function ($f_{\rm new}$, equation~\ref{eqn:fnew}) for a 2D disk, but it is not strictly self-consistent through Poisson's equation and the collisionless Boltzmann equation.
The underlying disk potential is set to,
\begin{equation}\label{eqn:Phi0}
	\Phi_0(R) = v_c^2 \ln(R/R_p),
\end{equation}
where the circular velocity is set so that $v_c=v_0$ and the scale length for the potential is $R_p=1$~kpc, in order to reproduce a flat rotation curve in two dimensions.  For the sake of analytic simplicity the spiral perturbation to the potential is assumed to be a density wave given by \citep{LYS69},
\begin{equation}
	\Phi_1(R,\phi,t) = \Phi_s(R,t) \cos\left[\alpha \ln(R/R_{\rm CR}) +m (\Omega_p t - \phi) \right],
\end{equation}
where $\phi$ is the azimuthal coordinate, and the constant $\alpha = m \cot \theta$ depends on the spiral pitch angle, $\theta$. 

The amplitude of the spiral perturbation,
\begin{equation}
	\Phi_s(R,t) = \dfrac{2\pi G R \Sigma(R) \epsilon(t)}{\alpha},
\end{equation}
depends on the radial surface density profile and the time-dependent fractional amplitude of the surface density, $\epsilon(t)$. 

The adopted time dependence for $\epsilon(t)$ has the Gaussian form,
\begin{equation}\label{eqn:et}
    \epsilon(t) = \epsilon_0\, e^{-(t-2T_{\rm Dyn})^2/2\sigma_t^2},
\end{equation}
so that the peak spiral amplitude occurs two orbital periods, $2T_{\rm Dyn}$, after the simulation begins, where the standard deviation is set to $\sigma_t=T_{\rm Dyn}$.  These assumptions ensure slow growth and decay for the perturbation, thus avoiding a non-adiabatic dynamical response. The value for the maximum fractional amplitude is set to $\epsilon_0=0.3$.

This study explores two forms for the radial surface density profile for the disk, $\Sigma(R)$.
The distribution function, $f_{\rm new}$, produces an exponential surface density profile in a 2D disk which is expressed analytically by,
\begin{equation}\label{eqn:expR}
    \Sigma(R) = \Sigma_0\, e^{-R/R_d}, 
\end{equation}
where the disk scale length for the surface density $R_d$ is set to match the prescription used for the distribution function $R_d = h_R = R_0/3$.  

An alternate set of models uses an inverse radial form for the surface density,
\begin{equation}\label{eqn:invR}
    \Sigma(R) = \dfrac{\Sigma_0}{R}.
\end{equation}
This form is self-consistent for the adopted \axisym\ potential, $\Phi_0(R)$ (equation~\ref{eqn:Phi0}), but not with the assumed distribution function (equation~\ref{eqn:fnew}).  
In all models $\Sigma_0$ is set so that the the surface density at $R_0$ is 50~M$_\odot$~pc$^{-2}$, similar to conditions in the solar neighbourhood of the Galaxy \citep[e.g.][]{KG91}.  

Test particle trajectories in each model are evolved for four orbital periods, where $T_{\rm Dyn}$ is given in Table~\ref{tbl:3models}.

Each model is named to signify its assumed radius of \CR, $R_{\rm CR}=\lbrace 4,5,6,7,8,9,10 \rbrace$~kpc, and adopted surface density profile.  For example, model M6e assumes $R_{\rm CR} = 6$~kpc and an exponential, \lq\lq e," surface density profile.  

\begin{deluxetable*}{lrrrrrrr}
	\tablecaption{Model Parameters \label{tbl:3models}}
	\tablehead{\colhead{Parameter} & \colhead{\hspace{0.8em}M4e \& M4i} & \colhead{\hspace{3.8em}M5e} & \colhead{\hspace{0.8em}M6e \& M6i} & \colhead{\hspace{3.8em}M7e} & \colhead{\hspace{0.8em}M8e \& M8i}& \colhead{\hspace{3.8em}M9e} &\colhead{\hspace{0.8em}M10e \& M10i}}
    \startdata
		$N$                & 25,000 & 15,000 & 50,000 & 25,000 & 25,000 & 15,000 & 25,000\\
        $R_{min}$ ($R_0$)  & 0.300  & 0.375  & 0.400  & 0.625  & 0.700  & 0.875  & 1.000\\
        $R_{\rm max}$ ($R_0$)  & 0.800  & 0.875  & 1.000  & 1.125  & 1.400  & 1.375  & 1.600\\  
        Area ($\pi R_0^2$) & 0.550  & 0.625  & 0.840  & 0.875  & 1.470  & 1.125  & 1.560 \\
		\hline
		$R_{\rm CR}$ (kpc)     & 4      & 5      & 6      & 7      & 8      & 9      & 10\\
		$T_{\rm Dyn}$ (Gyr)    & 0.11   & 0.14   & 0.17   & 0.20   & 0.22   & 0.25   & 0.28\\
	\enddata
    \vspace{1em}
	\tablecomments{Notes -- Parameters for orbital initial conditions (top section) are discussed in \S\ref{s:ICs}. Parameters for orbital integration (bottom section) are discussed in \S\ref{s:simulation}.  The annular area for the initial positions is also given in natural units.  Models using an exponential form ($\Sigma \propto e^{-R}$) for the surface density profile have lower-case \lq\lq e" at the end of the model name and models using an inverse radial form ($\Sigma \propto R^{-1}$) have a lower-case \lq\lq i."
	}
\end{deluxetable*}

\subsection{Resonant Trapping at Corotation}\label{s:trapping}

Stars in the plane of a non-\axisym\ potential have an invariant energy defined by the Jacobi integral,
\begin{equation}
	E_J = E - \Omega_p L_z,
\end{equation}
where the orbital energy, $E$, and angular momentum about the vertical axis, $L_z$, are taken to be in the non-rotating frame.  The orbital energy can be defined as,
\begin{equation}\label{eqn:Eran}
	E = E_c(R_L) + E_{\rm ran},
\end{equation}
where $E_c(R)$ is the orbital energy for a star in a circular orbit with radius $R$ in the unperturbed potential, $R_L$ is the radial coordinate in the unperturbed potential for a star in a circular orbit with unit mass angular momentum $L_z=R_L\, v_c$, and $E_{\rm ran}$ is the energy associated with non-circular motions, hereafter called \lq random energy'.  

Trapped orbits are stable orbits that are resonant with \CR\ (see discussion in \S\ref{s:theory}).  A star with zero random energy ($E_{\rm ran}=0$) is trapped when the parameter
\begin{equation}
	\Lambda_c = \dfrac{E_J - h_{\rm CR}}{|\Phi_s|_{\rm CR}}
\end{equation}
has absolute value equal to or less than unity ($-1 \leq \Lambda_c \leq 1$) \citep{Contopoulos78}, where the invariant $h_{\rm CR}$ is the Jacobi integral for a star in a circular orbit at the radius of \CR\ ($R_{\rm CR}$) in the unperturbed underlying \axisym\ potential.  For a star with some finite non-circular 
energy ($E_{\rm ran}$) to be trapped in a resonant orbit around \CR, the invariant $\Lambda_c$ must be replaced with the time-dependent quantity, $\Lambda_{\rm nc}(t)$ \citep[][their equation~22]{DW15}.   
For a rotationally supported disk, $v_{\rm ran}/v_c \lesssim 0.2$ where $v_{\rm ran}$ is the velocity associated with $E_{\rm ran}$, with a flat rotation curve, the value of $\Lambda_{\rm nc}(t)$ can be expressed as \citep[][their equation~33]{DW15},
\begin{equation}\label{eqn:Lambda_nc}
	\Lambda_{\rm nc}(t) = \Lambda_c - \left(\dfrac{R_L(t)}{R_{\rm CR}}\right)\left(\dfrac{E_{\rm ran}(t)}{|\Phi_s|_{\rm CR}}\right),
\end{equation}
and the criterion for resonant trapping around \CR\ is satisfied when,
\begin{equation}\label{eqn:capturecriterion}
	-1 \leq \Lambda_{\rm nc}(t) \leq 1.
\end{equation}
The expression in equation~\ref{eqn:Lambda_nc} assumes the \epi\ approximation in order to convert from action-space to energy-angular momentum space.

\begin{figure*}
	\includegraphics[width=\columnwidth]{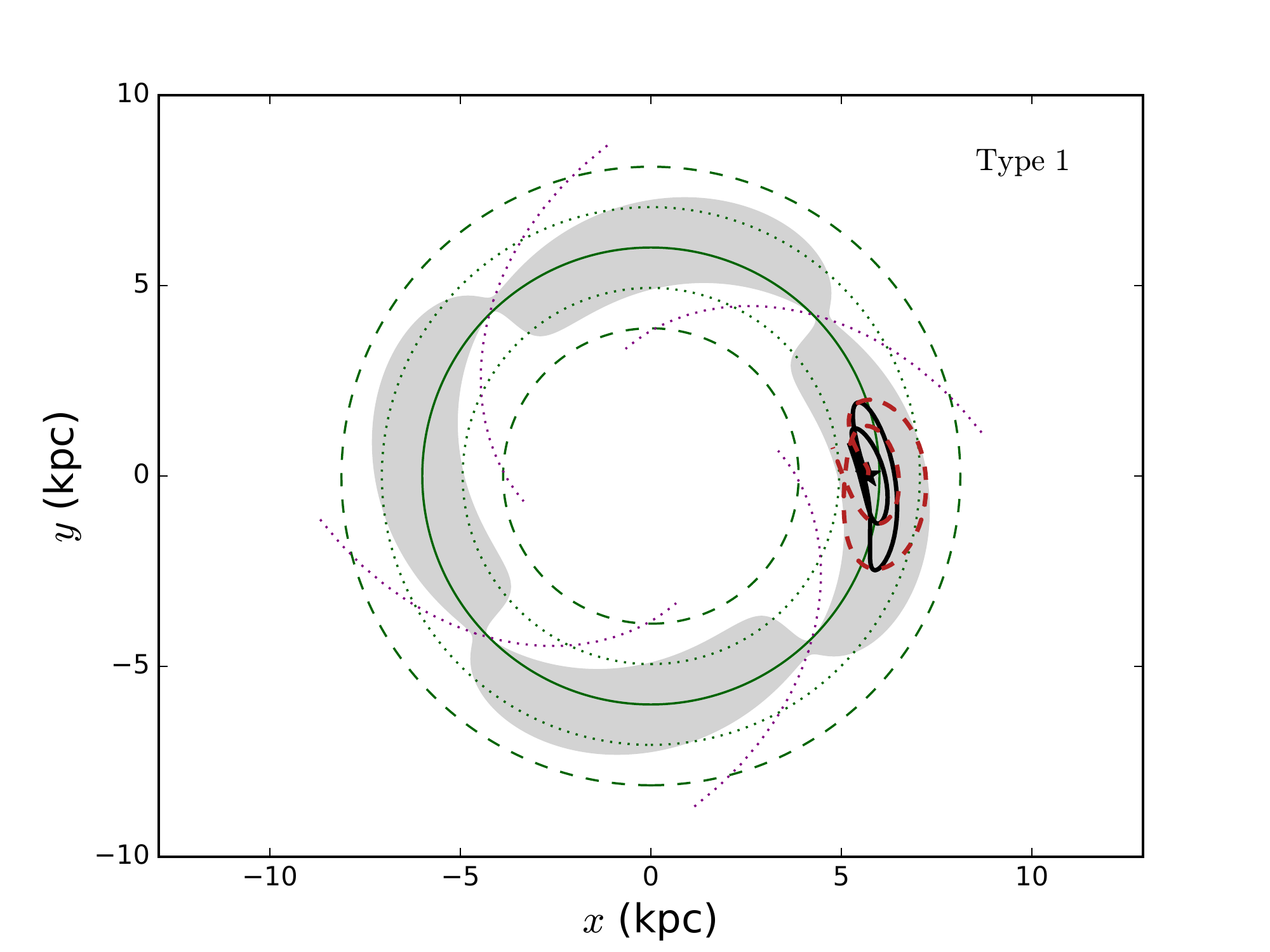}
	\includegraphics[width=\columnwidth]{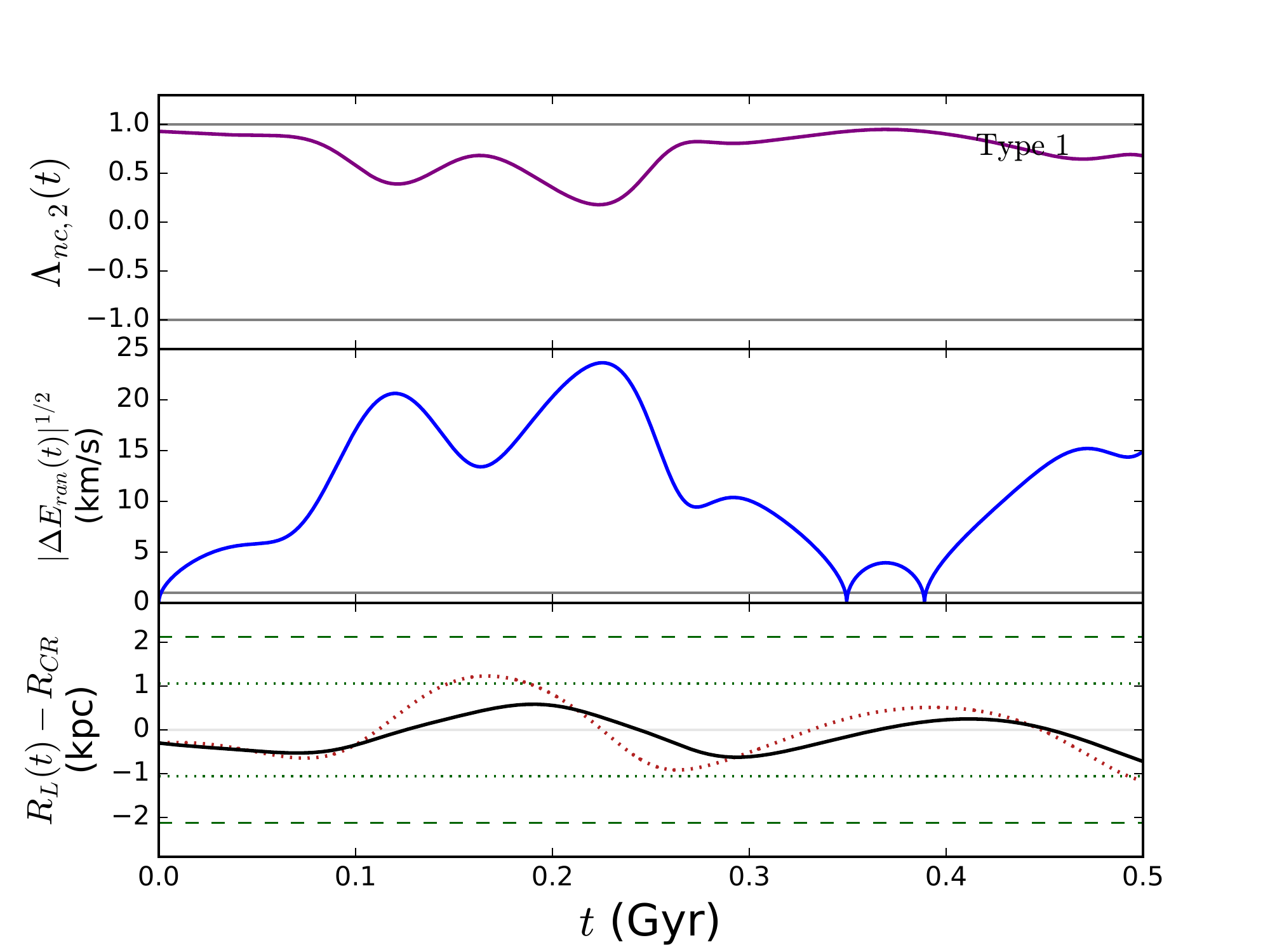}\\
    \caption{Illustrative example of a stable orbit trapped at the \CR\ resonance (Type~1). The potential is in Figure~\ref{fig:potentials_expR} for $R_{\rm CR}=6$~kpc.  This star particle is launched with initial 4D phase-space coordinates $(x,y,v_x,v_y)=(5.7$~kpc, $0$~kpc, $2$~km~s$^{-1}$, $220$~km~s$^{-1})$ (marked with a black star).  
    The fractional amplitude of the surface density is assumed to be $\epsilon = 0.3$ and the pitch angle $\theta = 30^\circ$.  The minima of the $m=4$ spiral pattern are indicated with dashed (magenta) curves.  Radius of \CR\ is shown as a solid (green) curve. Lindblad resonances are shown as long-dashed (green) curves and \ULR s are shown as dotted (green) curves.  
    The \CR\ region is shaded grey.  The the left hand panel shows the coordinate trajectory (dashed, red) and the guiding centre radius ($R_L$ -- solid, black) of the star particle.  The right hand panel shows the value for $\Lambda_{\rm nc}(t)$, which satisfies the capture criterion ($|\Lambda_{\rm nc}(t)|\leq 1$) at all times thus indicating it is in a trapped orbit at \CR. The middle right hand panel shows the square root of the absolute value for the change in energy associated with non-circular motions (\DelEran). This trapped orbit has variable circularity, but overall does not experience significant kinematic heating in a single trapped orbital period.  The bottom right hand panel shows the mean orbital radius (solid, black) -- and radial coordinate (dotted, red) -- oscillating about the \CR\,radius and never meets the \ULR\ resonance.}
    \label{fig:Type1_plots}
\end{figure*}

Figure~\ref{fig:Type1_plots} shows an example of an orbit that meets the criterion for a stable orbit trapped at the \CR\ resonance.  The left panel shows the trajectory (dashed, red) of the star particle evolved over 0.5~Gyr in a spiral potential with $R_{\rm CR}=6$~kpc. The mean orbital radius ($R_L$, equation~\ref{eqn:LR}) is indicated by the solid (black) curve and is in the \CR\ region (shaded) throughout the simulation.  
The radial coordinate (dotted, red), guiding centre radius (solid, black), their relative distances from the Lindblad resonances (dashed, green), and \ULR s (dotted, green) are also shown in the bottom of the right hand panel.  Since this is a trapped orbit the mean orbital radius radially oscillates about the \CR\ radius.
The top right hand panel of Figure~\ref{fig:Type1_plots} shows the value for $\Lambda_{\rm nc}(t)$ (equation~\ref{eqn:Lambda_nc}), which satisfies the \CR\ criterion (equation~\ref{eqn:capturecriterion}) at all times.   The middle right hand panel is square root of the absolute value for the change in energy associated with non-circular motions (\DelEran), which varies throughout the simulation, yet the orbit remains trapped with the \CR\ resonance.

\begin{figure}
	\includegraphics[width=0.5\columnwidth]{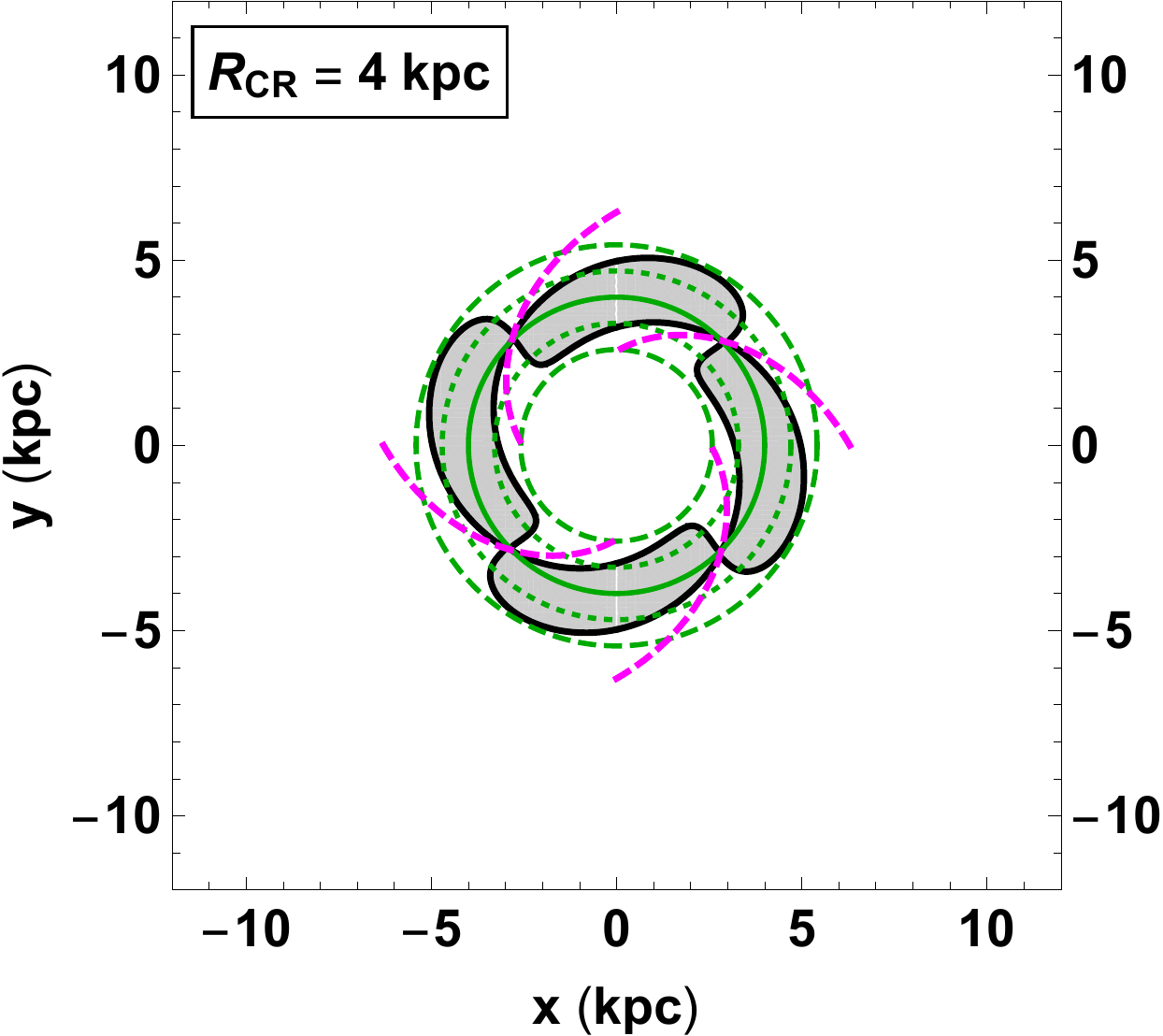}\includegraphics[width=0.5\columnwidth]{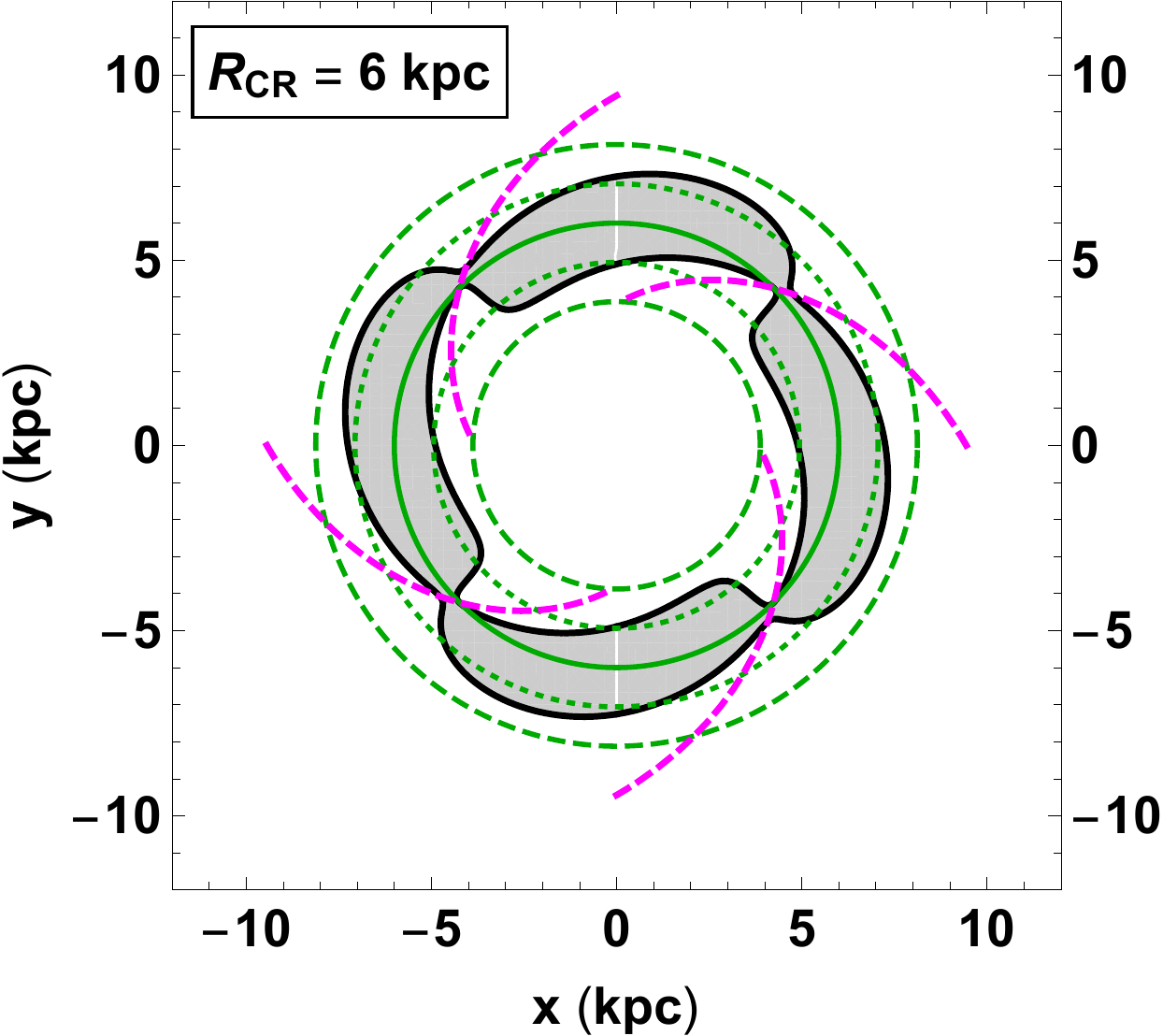}\\
    \includegraphics[width=0.5\columnwidth]{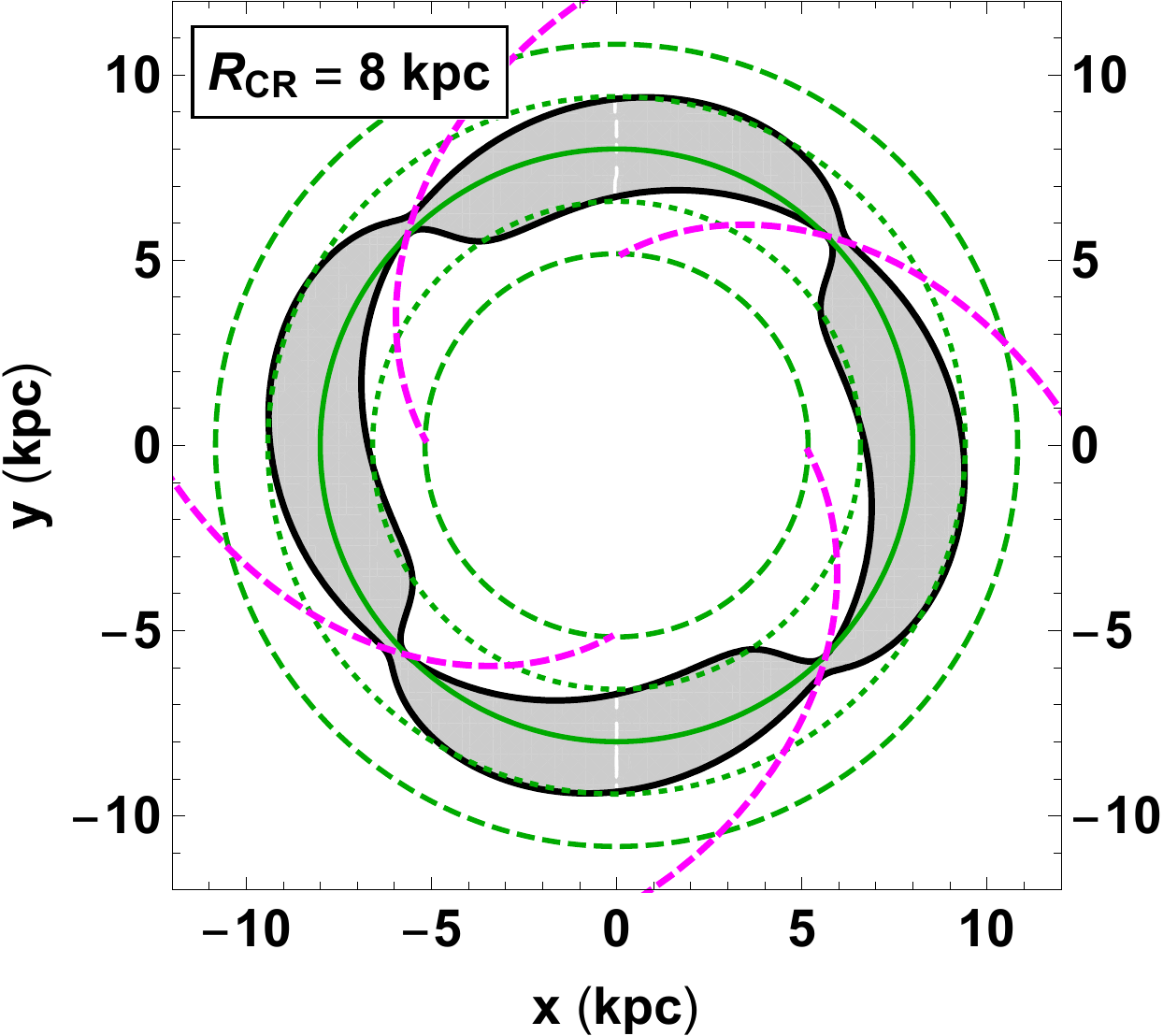}\includegraphics[width=0.5\columnwidth]{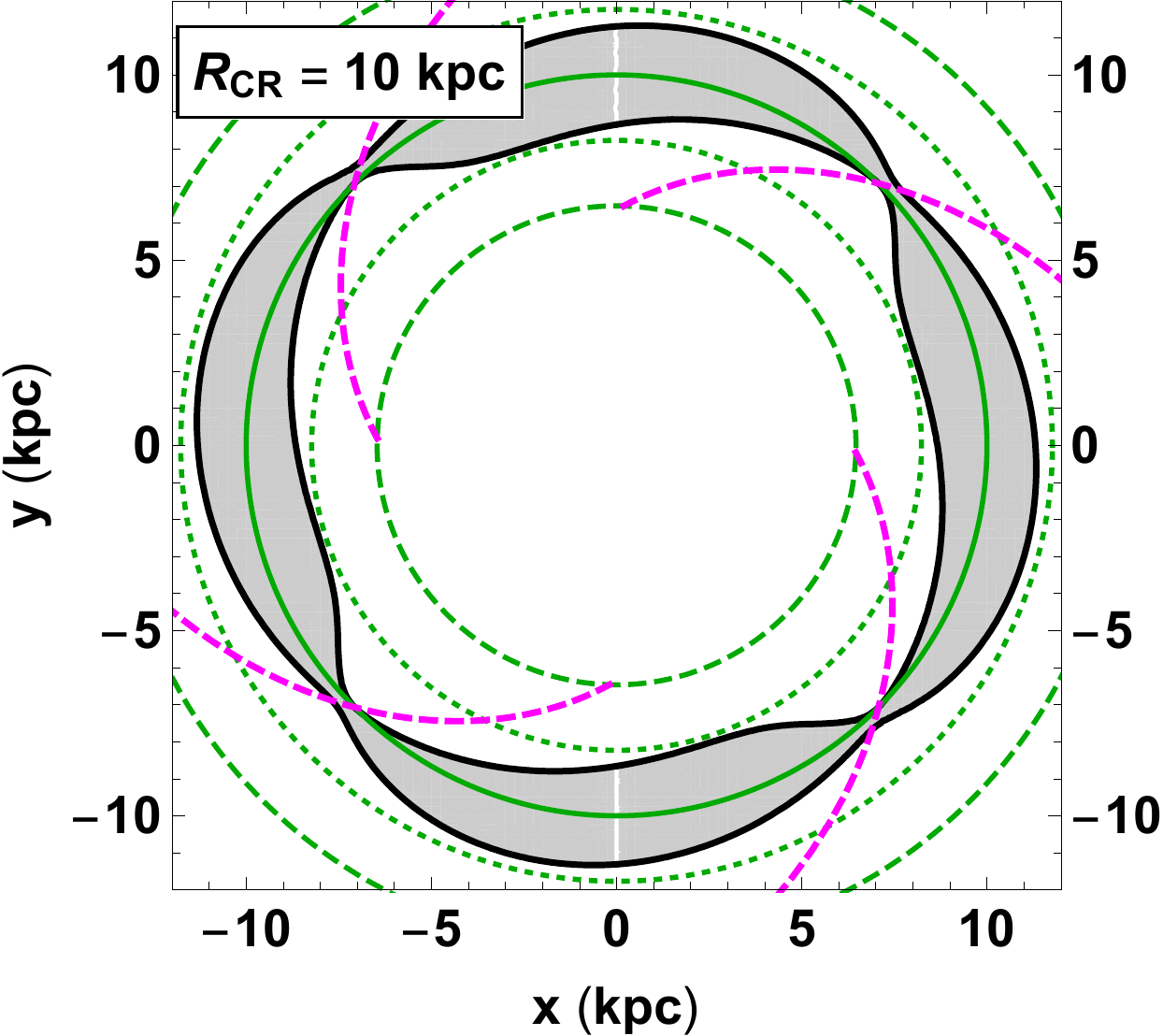}
    \caption{Contours showing the \CR\ region as a solid, thick (black) curve for models M4e, M6e, M8e, and M10e.  Radius of \CR\ for each model is given in the inset.
    Line-styles and colors of the curves in both the left hand panel and the bottom right hand panel match the definitions from Figure~\ref{fig:Type1_plots}.  
    The degree of resonant overlap depends on the radius of \CR\ for models assuming spiral amplitude scales with exponential surface density profile.}
    \label{fig:potentials_expR}
\end{figure}

\begin{figure}
	\includegraphics[width=0.5\columnwidth]{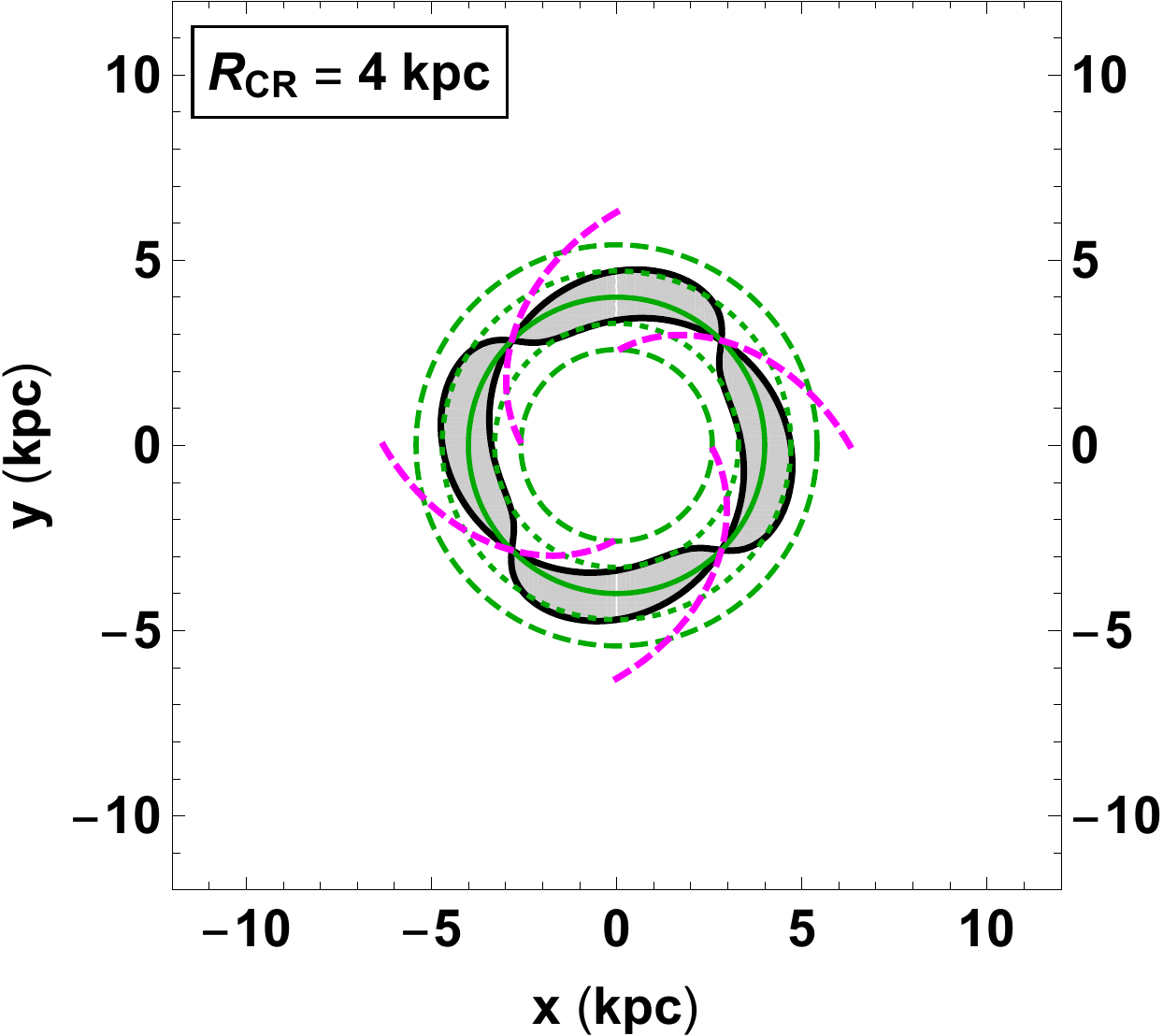}\includegraphics[width=0.5\columnwidth]{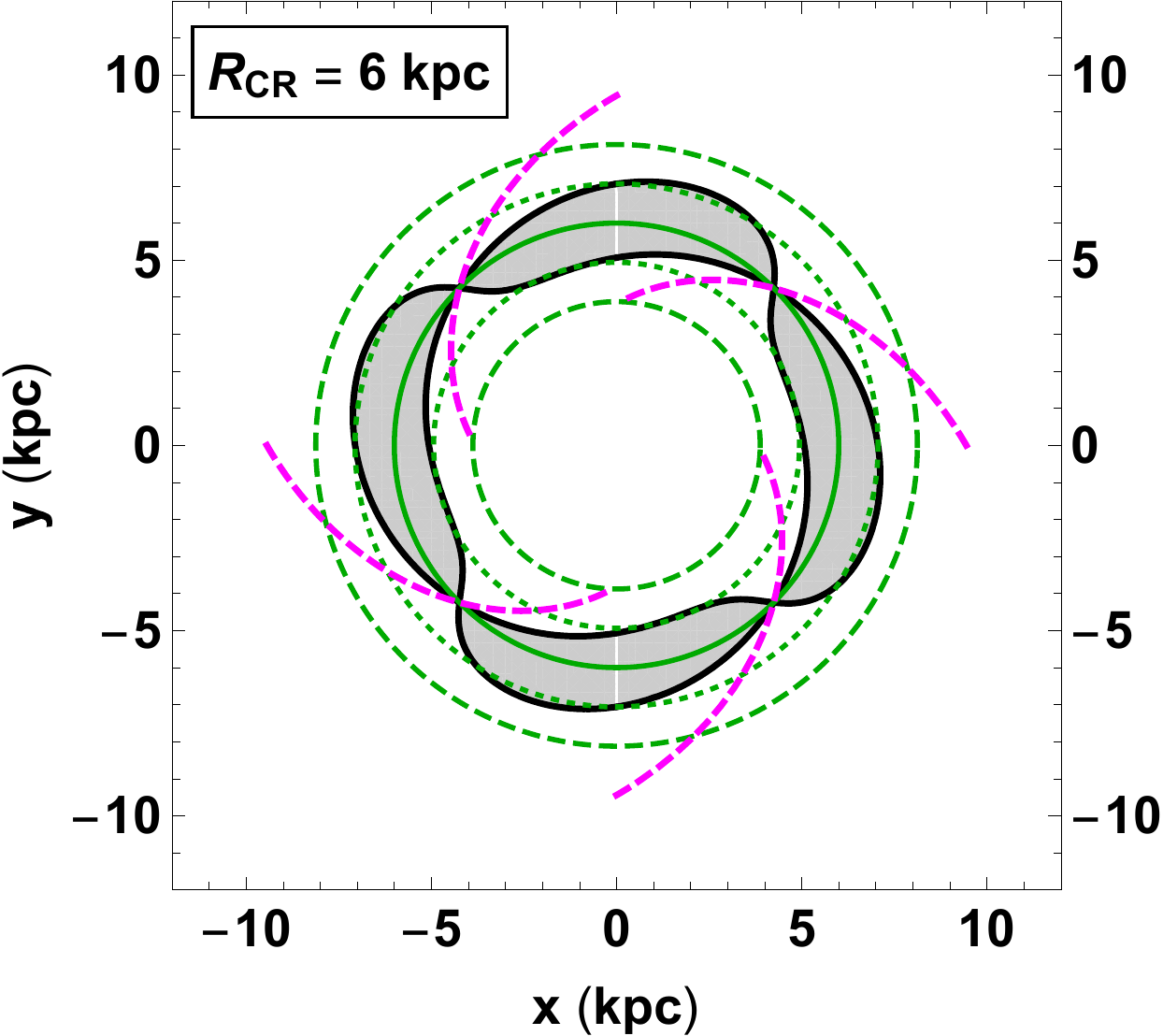}\\
    \includegraphics[width=0.5\columnwidth]{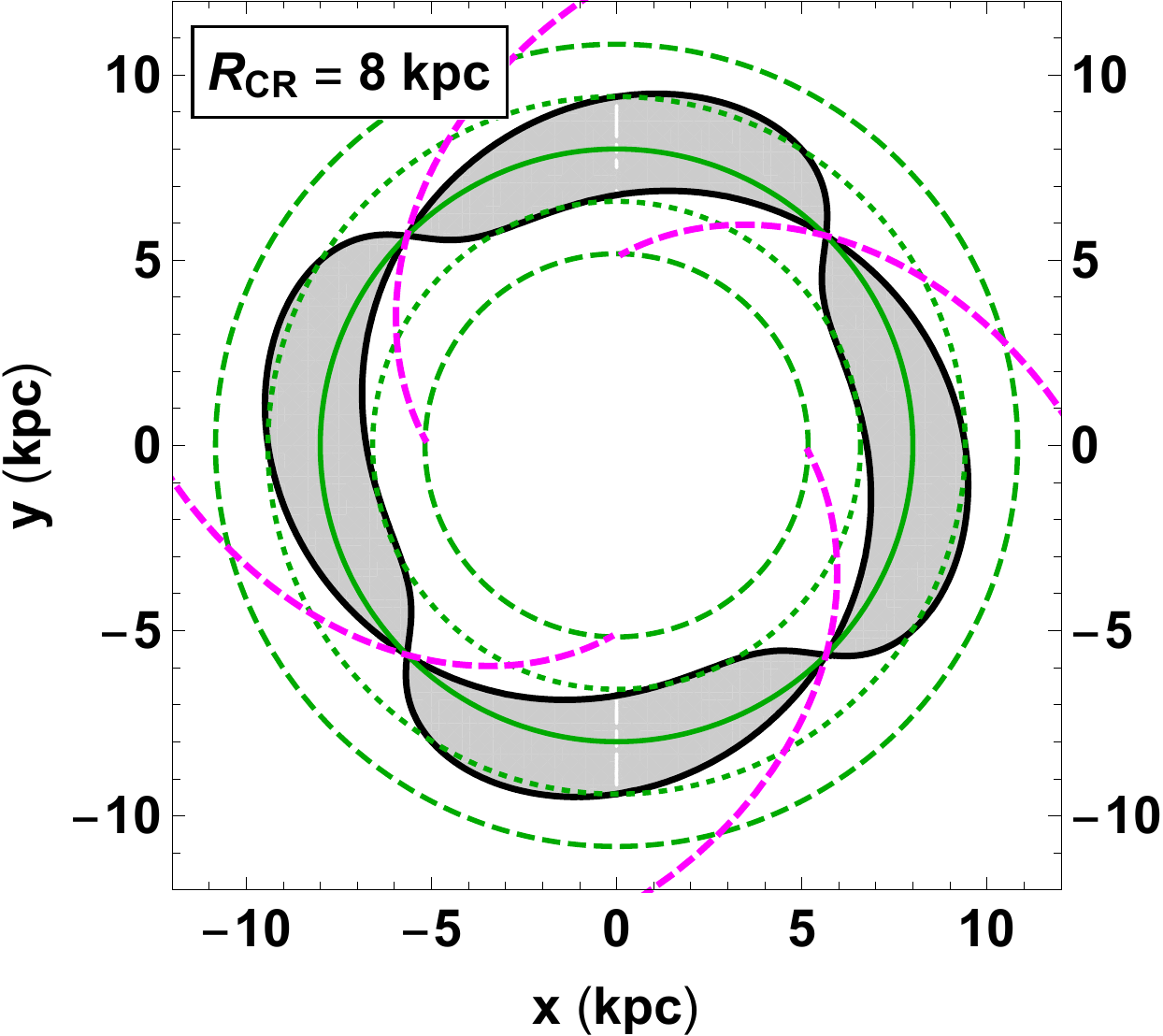}\includegraphics[width=0.5\columnwidth]{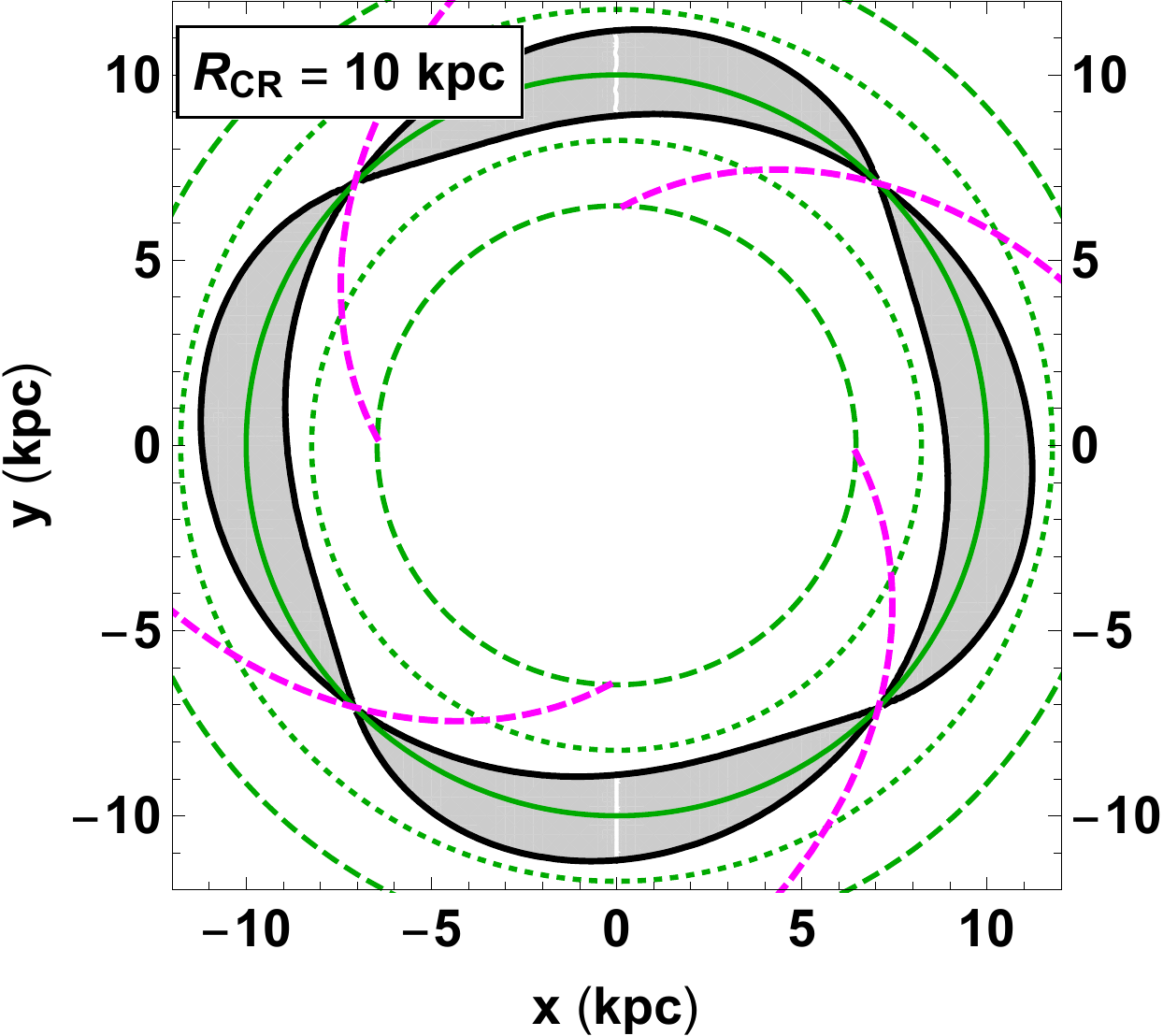}
    \caption{Contours showing the \CR\ region for models M4i, M6i, M8i, and M10i. Shading and linestyles have same meaning as in Figures~\ref{fig:Type1_plots} \&~\ref{fig:potentials_expR}. The degree of resonant overlap is nearly constant for models assuming a inverse radial surface density profile illustrating that the expected response for resonant overlap depends on the assumed model.}
    \label{fig:potentials_invR}
\end{figure}

Figures~\ref{fig:potentials_expR} and~\ref{fig:potentials_invR} show contours for the \CR\,region in models that use the exponential surface density profile in equation~\ref{eqn:expR} (M4e, M6e, M8e, and M10e) and the inverse radial surface density profile in equation~\ref{eqn:invR} (M4i, M6i, M8i, and M10i) for four different \CR\,radii.  These models were selected in order to illustrate how the distribution of resonances in the disk depends on the radius of \CR\,and assumed surface density profile when all other parameters are equal.  The \CR\,regions are shaded and outlined by a solid, thick (black) curve, where minima in the spiral potential are shown as short-dashed (magenta) curves.  Lindblad resonances are indicated by long-dashed (green) curves, \ULR s by dotted (green) curves, and radii of \CR\ by a solid (green) curves.  With decreasing \CR\ radius, the distances between the Lindblad resonances and \ULR s decrease while the area of the \CR\ region depends on the assumed surface density profile.  For models assuming an exponential disk, the \CR\,region is broader toward the galactic center and therefore trends toward a greater degree of resonant overlap for spiral patterns with a smaller \CR\,radius (and higher pattern speed).  Models adopting the inverse radial surface density profile have radially independent spiral strength and therefore the degree of resonant overlap depends only on the fractional amplitude for the spiral pattern for constant number of arms and pitch angle.\\

\subsection{Orbital categorization Scheme}\label{s:classification}

\begin{deluxetable}{ll}
	\tablecaption{Orbital Categorization Scheme \label{tbl:orbitcat}}
	\tablehead{\colhead{Category} & \colhead{Description}}
	\startdata
		Type~1 & Trapped in resonant orbit about \CR\vspace{0.5em}\\
        Type~2 & Never trapped at \CR \vspace{0.5em}\\
        Type~3$\rightarrow$1 & In a trapped orbit before \& after crossing \\
		         & an \ULR \vspace{0.5em}\\
		Type~3$\rightarrow$2 & In a trapped before crossing an \ULR,\\
                 & but not after \vspace{0.5em}\\
		Other & Any other scenario\\
	\enddata
\end{deluxetable}

The expectation for a star trapped at the \CR\ resonance is that its mean orbital radius ($R_L$) will oscillate about \CR\ ($R_{\rm CR}$) indefinitely with negligible change to its circularity.  The primary focus of this work is to critically examine that expectation in cases when a trapped star simultaneously meets another resonant criterion.

\begin{figure*}
    \includegraphics[width=\columnwidth]{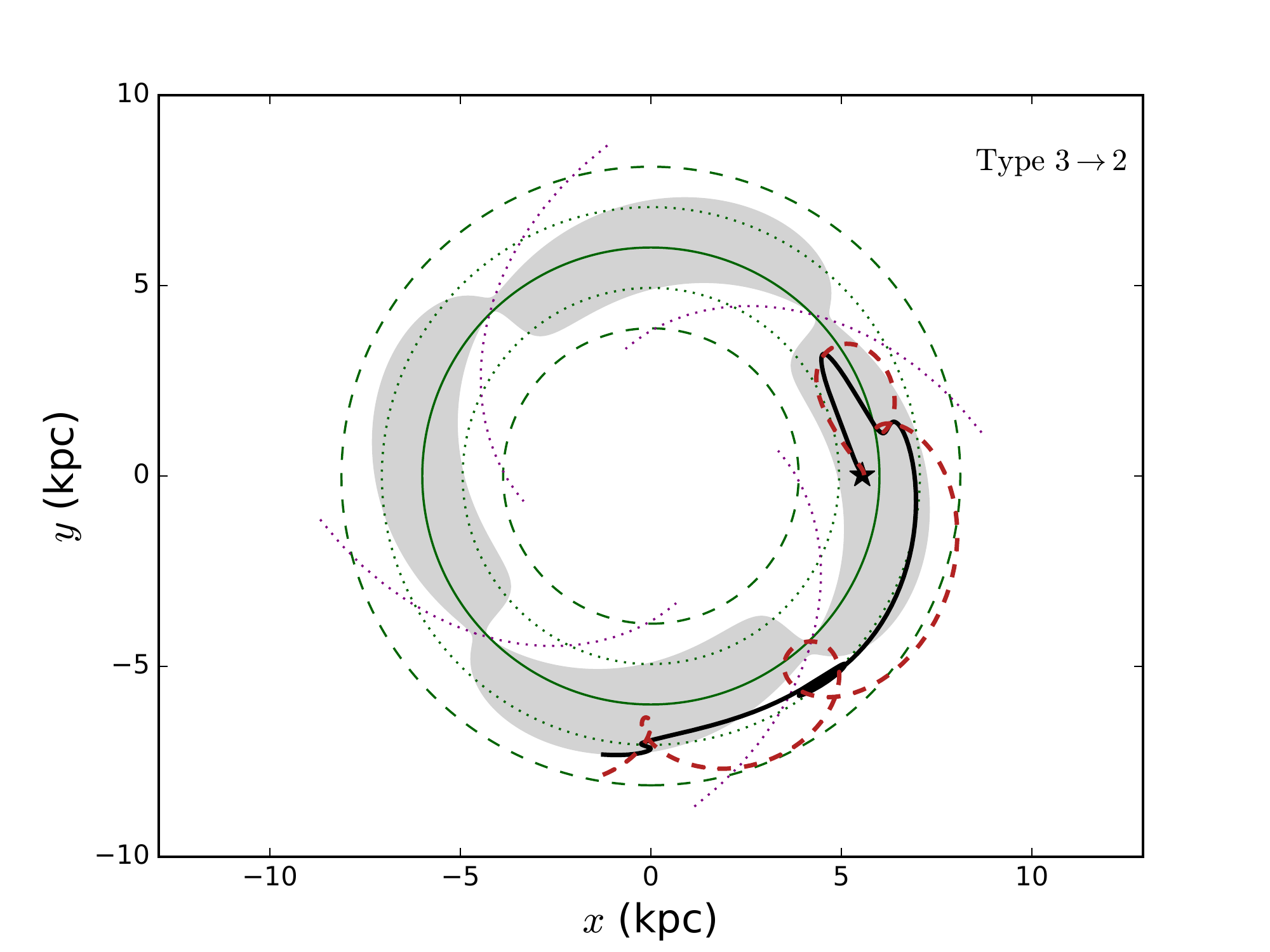}
    \includegraphics[width=\columnwidth]{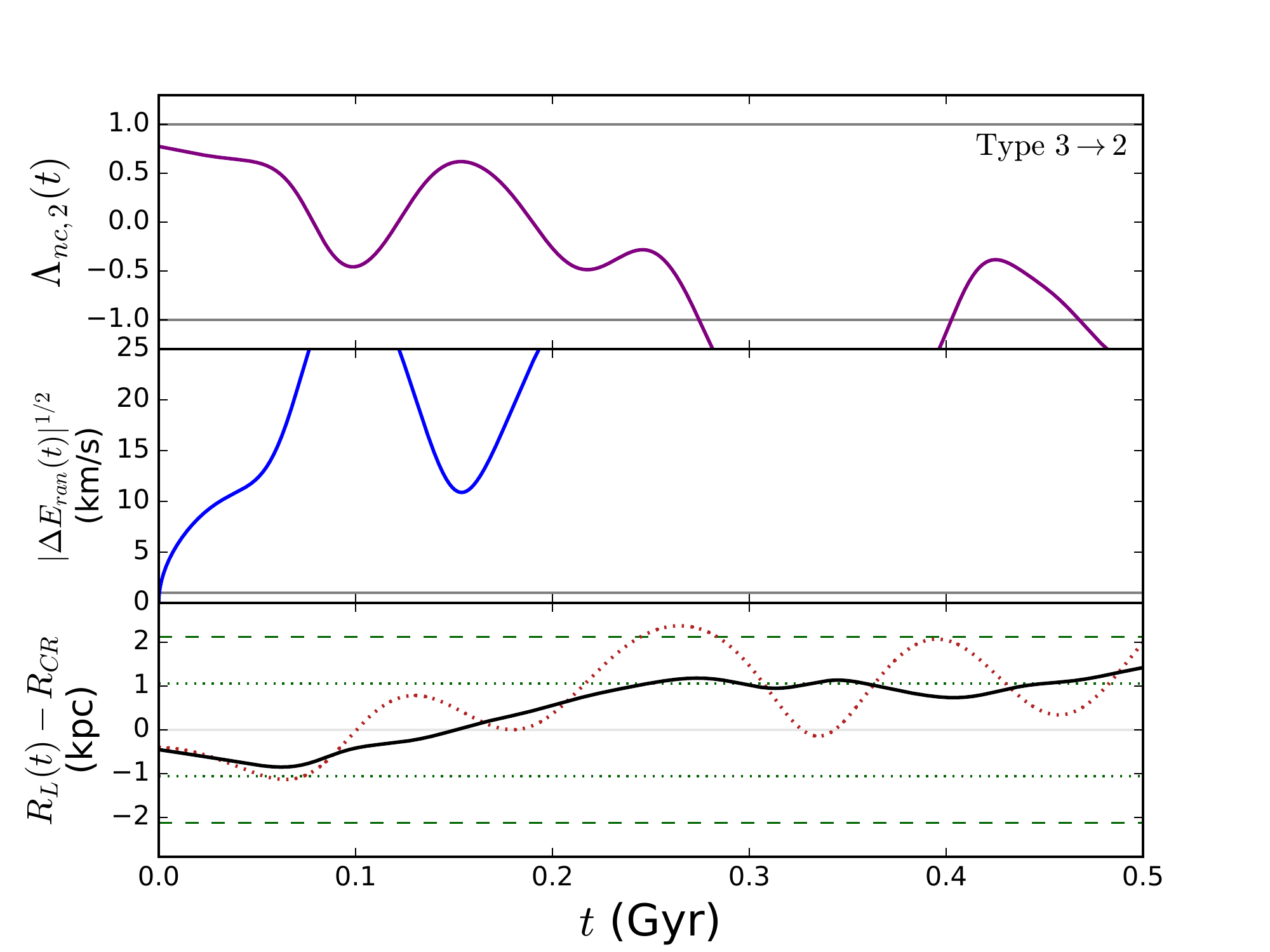}\\
    \caption{Illustrative example of the orbital response for a star that is initially in a stable orbit trapped at the \CR\ resonance that later meets a second resonant criterion when the guiding centre radius crosses an \ULR\ (Type~3).  Initial 4D phase-space coordinates for this star are $(x,y,v_x,v_y)=(5.6$~kpc, $0$~kpc, $-1$~km~s$^{-1}$, $218$~km~s$^{-1})$.  The line-styles and shading have the same meaning as in Figure~\ref{fig:Type1_plots}.   Once the star particle crosses the \ULR\,it is in resonant overlap for nearly 0.25~Gyr.  During this time the star particle is \kmy\ heated, is not trapped as it crosses a spiral arm and then becomes temporarily trapped before ending in a non-resonant orbit (Type~3$\rightarrow$2).}
    \label{fig:Type32_plots}
\end{figure*}

Each test particle's trajectory is used to calculate a time series array for its orbital angular momentum ($L_z$), the associated mean orbital radius ($R_L$), the random orbital energy ($E_{\rm ran}$), and $\Lambda_{\rm nc}(t)$.  These arrays give a quantitative measure for whether or not a star instantaneously meets any dynamical resonant criteria.  A particle is trapped in a resonant orbit about \CR\ when equation~\ref{eqn:capturecriterion} is satisfied via the instantaneous value for $\Lambda_{\rm nc}(t)$ (equation~\ref{eqn:Lambda_nc}).  It is resonant with a Lindblad or \ULR\ resonance when $R_L(t) = R_{\rm LR}^{(n+1)}$ (equation~\ref{eqn:LR}).

Each of these quantities is calculated using a the time-dependent spiral potential described in \S\ref{s:simulation} with the exception of the value for $\Lambda_{\rm nc}(t)$.  The value for $\Lambda_{\rm nc}(t)$ is calculated using a time-independent spiral amplitude set to equal the time-dependent potential's maximum amplitude, $\Phi_s(R)=\Phi_s(R,2T_{\rm Dyn})$.  
This choice was made for the following reason.  In a potential that includes a transient spiral pattern, all stars begin and end in orbits that are not trapped at \CR\,since \CR\,only exists in the presence of a non-\axisym\,perturbation.  Using constant $\Phi_s(R)$ to evaluate $\Lambda_{\rm nc}(t)$ identifies stars that would remain trapped for the duration of the simulation when the spiral amplitude is large enough for them to be trapped.  
By adopting this definition, it is possible to distinguish between star particles that are no longer trapped after an episode of resonant overlap from those that are temporarily trapped in Type~1 orbits due to spiral amplitude growth and decay.  A comparative study showed that by using a time-dependent spiral potential to calculate $\Lambda_{\rm nc,2}(t)$ there was a small degree of contamination of Type~4 orbits with visually identified Type~1 orbits.

Each trajectory is categorized into one of five types.  Star particles that remain in a stable, resonant orbit about \CR\ for the duration of the simulation are categorized as Type~1.  Type~2 orbits are those that never satisfy this criterion and are therefore never in trapped orbits, circulating about the galactic centre in the frame rotating with the spiral pattern for the duration of the simulation.  Type~3 orbits begin in trapped orbits and have at least one episode when the mean orbital radius crosses an \ULR, thus simultaneously meeting two dynamical resonant criteria.  Type~3 orbits are subdivided into Type~3$\rightarrow$1, which remain in trapped orbits after crossing an \ULR, and Type~3$\rightarrow$2, which are not in trapped orbits at the end of the simulation after crossing an \ULR.  All other cases are categorized as \lq Other'.  In no case does a trapped orbit have a mean orbital radius that crosses a Lindblad resonance.  A summary of these categories is given in Table~\ref{tbl:orbitcat}.

Figure~\ref{fig:Type32_plots} shows an illustrative example of a Type~3$\rightarrow$2 orbit, where the star particle does not remain in a trapped orbit after crossing the outer \ULR.  In this example, the orbital trajectory is evolved over 0.5~Gyr and begins in a trapped orbit.  At $t\sim 0.25$~Gyr, the star particle is in resonance with both \CR\ ($1\leq|\Lambda_{\rm nc,2}|$) and the outer \ULR\ ($R_L=R^{(2)}_{\rm LR}$) resulting in an increase in non-circular orbital energy ($E_{\rm ran}$).  This kinematic heating causes the orbit to no longer be resonant with \CR\ ($\Lambda_{\rm nc,2}<-1$) and the star particle is therefore able to cross a spiral arm.  The star is briefly trapped starting at $t\sim 0.4$~Gyr, but ends in an orbit that is resonant with the outer \ULR\ only.  The orbital response to resonant overlap, being chaotic, is highly variable for very similar initial conditions.  In practice, this means that Type~3 orbits commonly have intervals when the orbital trajectory oscillates between being trapped and not trapped with the \CR\ resonance before being permanently non-trapped.  As such, it is likely that the distinction between Type~3$\rightarrow$1 and Type~3$\rightarrow$2 is somewhat unresolved since it is a time-dependent and, given a long enough time-scale, many Type~3$\rightarrow$1 orbits eventually evolve into Type~3$\rightarrow$2 orbits. However, the time-scale for the transient spiral arms in the current simulations ($\sigma_t=T_{\rm Dyn}$) is several times less than the time-scale for the simulations ($T=4\,T_{\rm Dyn}$) and so the classifications presented here are expected to be qualitatively robust. We have not accounted for changes to the spiral perturbation's pattern speed.  Analyses of the pattern speeds in simulations with transient spiral arms typically show a time-dependence for spiral pattern speed \citep[e.g.,][]{Roskar12,dOnghia16}.  Further exploration is required in order to understand how an evolving pattern speed would affect trapped orbits.

\begin{rotatetable*}
\begin{deluxetable*}{lrrrrrrr}
	\tablecaption{Distribution of Orbits for Exponential Surface Density Profile\label{tbl:norbit_expR}}
    \tablehead{\colhead{ }& \colhead{M4e} & \colhead{M5e} & \colhead{M6e} & \colhead{M7e} & \colhead{M8e} & \colhead{M9e} & \colhead{M10e}}
    \startdata
		Type~1                        & 842 ($3.4\%$)     & 862 ($5.8\%$)    & 3,910 ($7.8\%$)   & 2,616 ($17.4\%$) & 4,782 ($19.1\%$)  & 4,999 ($33.3\%$) & 8,799 ($35.2\%$) \\
        Type~2                        & 12,761 ($51.0\%$) & 7,293 ($48.6\%$) & 29,179 ($58.4\%$) & 6,631 ($44.2\%$) & 14,306 ($57.2\%$) & 6,861($45.7\%$)  & 13,080 ($52.3\%$) \\
        Type~3$\rightarrow$1          & 3,027 ($12.1\%$)  & 1,794 ($12.0\%$) & 4,547 ($9.1\%$)   & 1,169 ($7.8\%$)  & 952 ($3.8\%$)     & 463 ($3.1\%$)    & 314 ($1.3\%$) \\
		Type~3$\rightarrow$2          & 4,462 ($17.9\%$)  & 2,733 ($18.2\%$) & 6,357 ($12.7\%$)  & 2,322 ($15.5\%$) & 2,110 ($8.4\%$)   & 740 ($4.9\%$)    & 265 ($1.1\%$)  \\
		Other                         & 3,908 ($15.6\%$)  & 2,318 ($15.5\%$) & 6,007 ($12.0\%$)  & 2,262 ($15.1\%$) & 2,849 ($11.4\%$)  & 1,937 ($12.9\%$) & 2,542 ($10.2\%$) \\
       	\hline
        Type~3                        & 7,489 ($30.0\%$)  & 4,527 ($30.2\%$) & 10,904 ($21.8\%$) & 3,491 ($23.3\%$) & 3,062 ($12.3\%$)  & 1,203 ($8.0\%$)  & 579 ($2.3\%$)  \\
        Type~1+3                      & 8,331 ($33.3\%$)  & 5,389 ($35.9\%$) & 14,814 ($29.6\%$) & 6,107 ($40.7\%$) & 7,844 ($31.4\%$)  & 6,202 ($41.4\%$) & 9,378 ($37.5\%$)  \\
        \hline
        Type~3$\rightarrow$2/Type~3   & $59.6\%$          & $60.4\%$         & $58.3\%$          & $66.5\%$         & $68.9\%$          & $61.5\%$         & $45.8\%$  \\
        Type~3/Type~1+3               & $89.9\%$          & $84.0\%$         & $73.6\%$          & $57.2\%$         & $39.0\%$          & $19.4\%$         & $6.2\%$ \\
        Type~3$\rightarrow$2/Type~1+3 & $53.6\%$          & $50.7\%$         & $42.9\%$          & $38.0\%$         & $26.9\%$          & $11.9\%$         & $2.8\%$ \\
    \enddata    
    \vspace{1em}
	\tablecomments{Notes -- Number and percentage of trajectories that satisfy the orbital categorization scheme outlined in Table~\ref{tbl:orbitcat} for models that use an exponential surface density profile ($\Sigma(R)\propto e^{-R}$).}
\end{deluxetable*}
\end{rotatetable*}

The top panel in Table~\ref{tbl:norbit_expR} and Table~\ref{tbl:norbit_invR} summarizes the distribution of orbits among each of the orbital categories for each model.  

The middle panel in each table gives the total the number and percentage of orbits that experience resonant overlap (Type~3) in each simulation as well as the number and percentage of all orbits that have initial conditions for trapped orbits (Type~1+3).  The trend in the initial fraction of trapped orbits roughly follows the ratio between the area of the \CR\,region and the area of the \anu\,of sampled initial conditions for each model. \citep[Also see Figures~11 \&~14d for Model~W -- and relevant discussion -- in][which show the maximum width for the \CR\ region and fraction of stars initially in trapped orbits for  model parameters equivalent to those in Table~\ref{tbl:norbit_expR}.]{DW18}  

\begin{figure*}
    \includegraphics[width=\columnwidth]{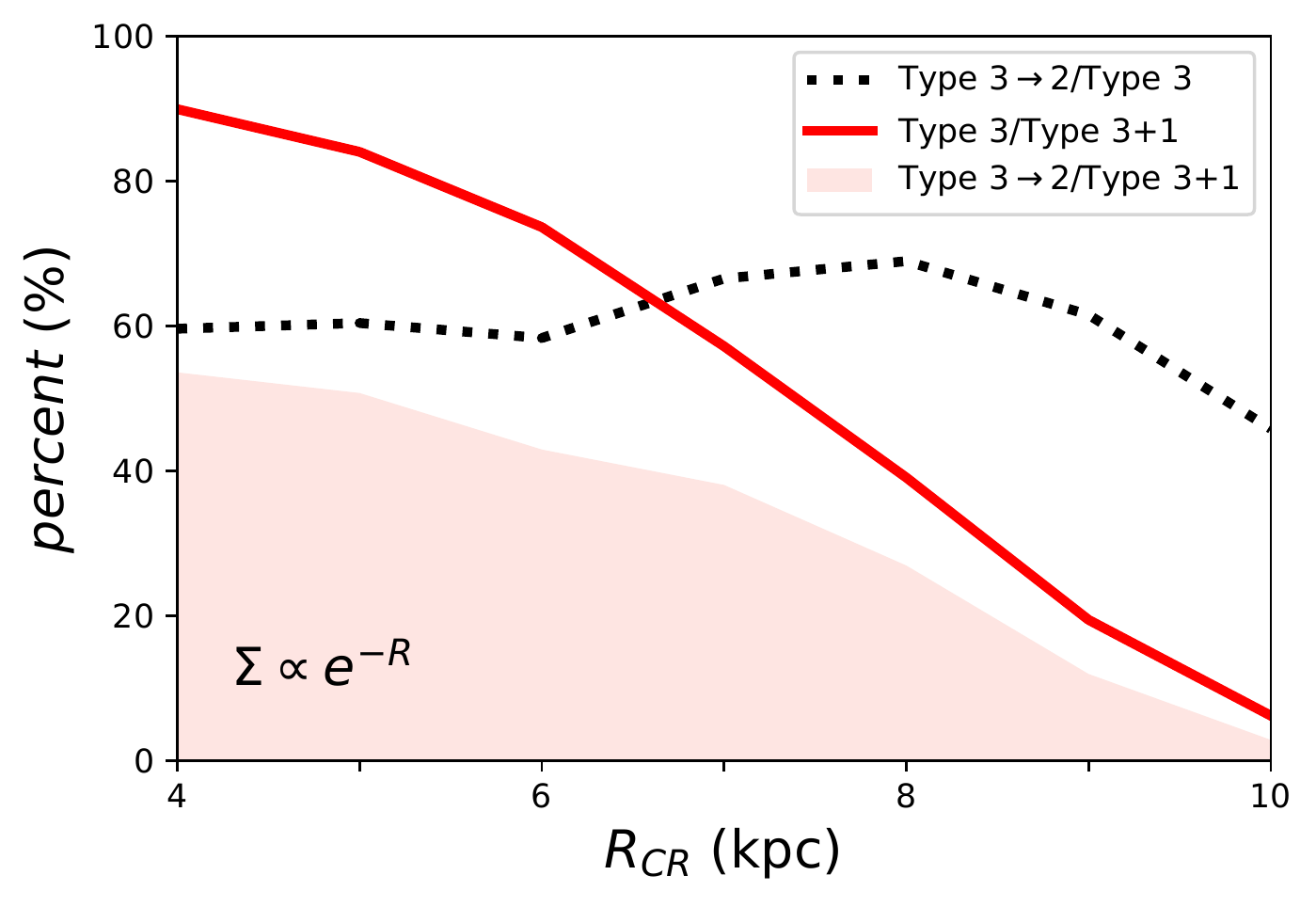}
    \includegraphics[width=\columnwidth]{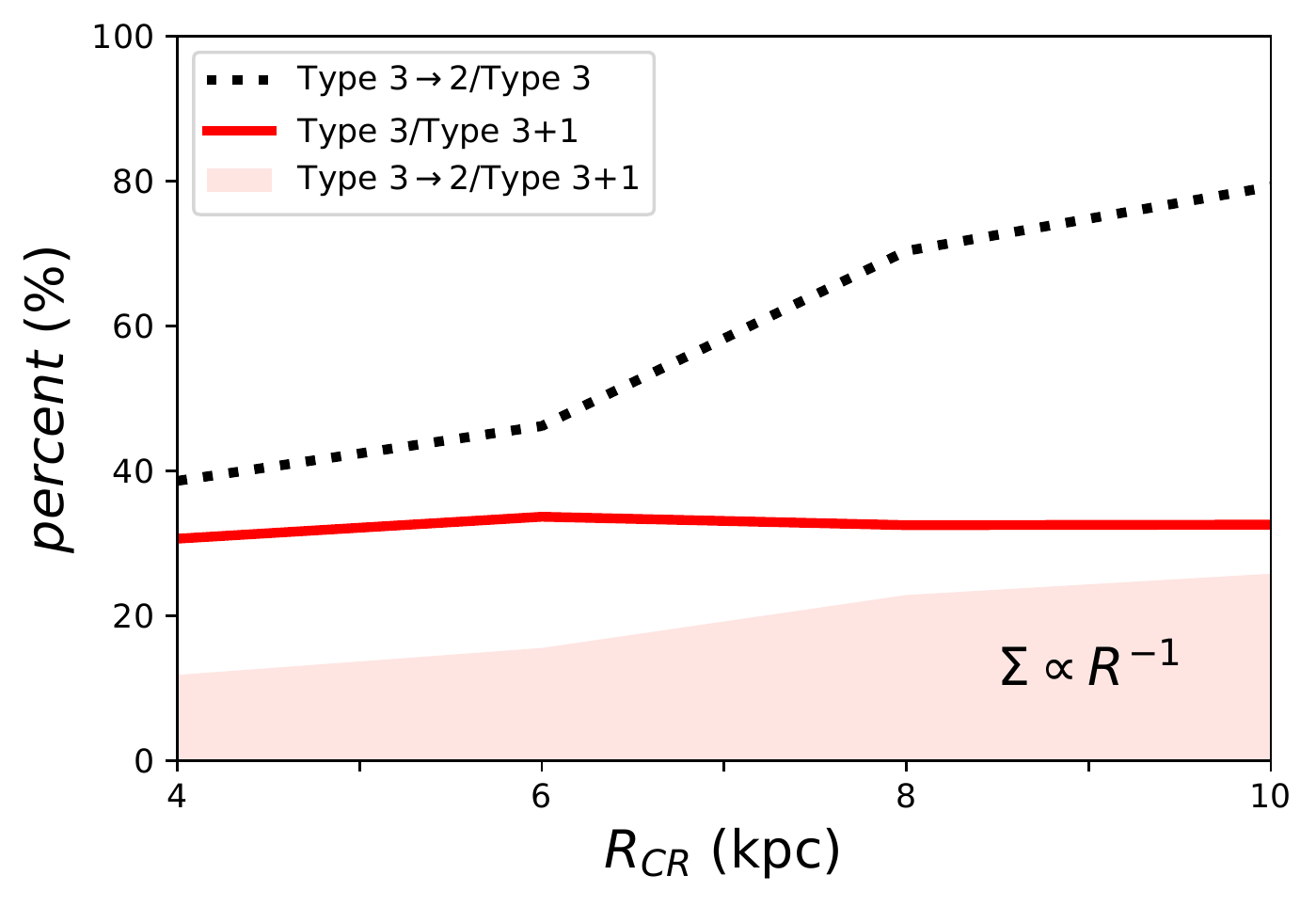}\\
    \caption{Curves showing the fractional percent for three physically notable orbital classifications.  The left plot is for models using an exponential surface density profile ($\Sigma(R)\propto e^{-R}$) to determine spiral strength and the right plot is for models using an inverse radial profile ($\Sigma(R)\propto R^{-1}$).  The fraction of trapped orbits that experience resonant overlap with an \ULR\,is indicated with a solid (red) line (Type~3/Type~3+1).  The fraction of those orbits which experience resonant overlap and that are permanently not trapped after crossing an \ULR\,are indicated with a dotted (black) line (Type~3$\rightarrow$2/Type~3).  The shaded (salmon) region shows the fraction of all trapped orbits that are no longer in trapped orbits after an episode of resonant overlap (Type~3$\rightarrow$2/Type~3+1).}
    \label{fig:4b}
\end{figure*}

The bottom panel in each table quantifies several scaling relations that are illustrated in Figure~\ref{fig:4b}.
In all models, between ~40-80\% of trapped stars have an episode of resonant overlap and are not in trapped orbits at the end of the simulation (Type~3$\rightarrow$2/Type~3) as indicated by a dotted (black) line in Figure~\ref{fig:4b}. This suggests that the dynamical response of trapped stars to resonant overlap alters a significant number of orbital classifications from those that lead to well behaved \RM\,due to \ct\ to orbits that inhabit other regions of the phase-space and could be \kmy\ heated.
The fraction of the orbits that are initially trapped (Type~3+1) and experience resonant overlap (Type~3) during the total time evolved ($4T_{\rm Dyn}$) is indicated by a solid (red) line (Type~3/Type~3+1).  For the series of models that uses an inverse radial surface density profile ($\Sigma\propto R^{-1}$), the radial size of the \CR\,region closely scales with the distance between the \ULR s ($\Delta R_{\rm LR}^{(n+1)}$), so the degree of resonant overlap is nearly constant for these models ($\sim 30\%$).
For the series of models that uses an exponential surface density profile ($\Sigma\propto e^{-R}$) the degree of resonant overlap increases toward the galactic center from $\sim 6\%$ for M10e to $\sim 90\%$ for M4e.  
The shaded (salmon) region shows the ratio of the number of stars initially in trapped orbits that experience resonant overlap and end the simulation in non-trapped orbits to the number of stars that begin in trapped orbits (Type~3$\rightarrow$2/Type~1+3).  These trends are similar to those for Type~3/Type~1+3 since the fraction for Type~3$\rightarrow$2/Type~3 is nearly constant. 

\begin{deluxetable*}{lrrrr}
	\tablecaption{Distribution of Orbits for Inverse Radial Surface Density Profile\label{tbl:norbit_invR}}
    \tablehead{	\colhead{ }  &\colhead{M4i} & \colhead{M6i} & \colhead{M8i} & \colhead{M10i}} 
    \startdata
		Type~1                        & 4,598 ($18.4\%$)  & 8,968 ($17.9\%$)  & 5,199 ($20.8\%$)  & 5,062 ($20.3\%$) \\
        Type~2                        & 12,596 ($50.4\%$) & 29,074 ($58.2\%$) & 14,415 ($57.7\%$) & 13,195 ($52.8\%$) \\
        Type~3$\rightarrow$1          & 1,245 ($5.0\%$)   & 2,447 ($4.9\%$)   & 742 ($3.0\%$)     & 508 ($2.0\%$)  \\
		Type~3$\rightarrow$2          & 782 ($3.1\%$)     & 2,097 ($4.2\%$)   & 1,758 ($7.0\%$)   & 1,934 ($7.7\%$) \\
		Other                         & 5,779 ($23.1\%$)  & 7,414 ($14.8\%$)  & 2,886 ($11.5\%$)  & 4,301 ($17.2\%$) \\
       	\hline
        Type~3                        & 2,027 ($8.1\%$)   & 4,544 ($9.1\%$)   & 2,500 ($10.0\%$)  & 2,442 ($9.8\%$) \\
        Type~1+3                      & 6,625 ($26.5\%$)  & 13,512 ($27.0\%$) & 7,699 ($30.8\%$)  & 7,504 ($30.0\%$) \\
        \hline
        Type~3$\rightarrow$2/Type~3   & $38.6\%$          & $46.2\%$         & $70.3\%$           & $79.2\%$         \\
        Type~3/Type~1+3               & $30.6\%$          & $33.6\%$         & $32.5\%$           & $32.5\%$         \\
        Type~3$\rightarrow$2/Type~1+3 & $11.8\%$          & $15.5\%$         & $22.8\%$           & $25.8\%$         \\
    \enddata    
    \vspace{1em}
	\tablecomments{Notes -- Number and percentage of trajectories that satisfy the orbital categorization scheme outlined in Table~\ref{tbl:orbitcat} for models that use the inverse radial surface density profile ($\Sigma(R)\propto R^{-1}$).}
\end{deluxetable*}

\section{discussion}\label{sec:discussion}

It is expected that \RM\,of stars as a result of \ct\,leads to a redistribution of orbital angular momentum with no associated kinematic heating \citep{SB02}.  The following discussion first suggests a revision to the assumption that \ct\,from a single spiral pattern is necessarily a cold process (\S\ref{s:heating}).  It then considers  limits to the radial range within which \ct\,is \kmy\,cold (\S\ref{s:DeltaRmax}).

\subsection{Kinematic heating from \ct}\label{s:heating}

\begin{figure*}
    \includegraphics[width=2\columnwidth]{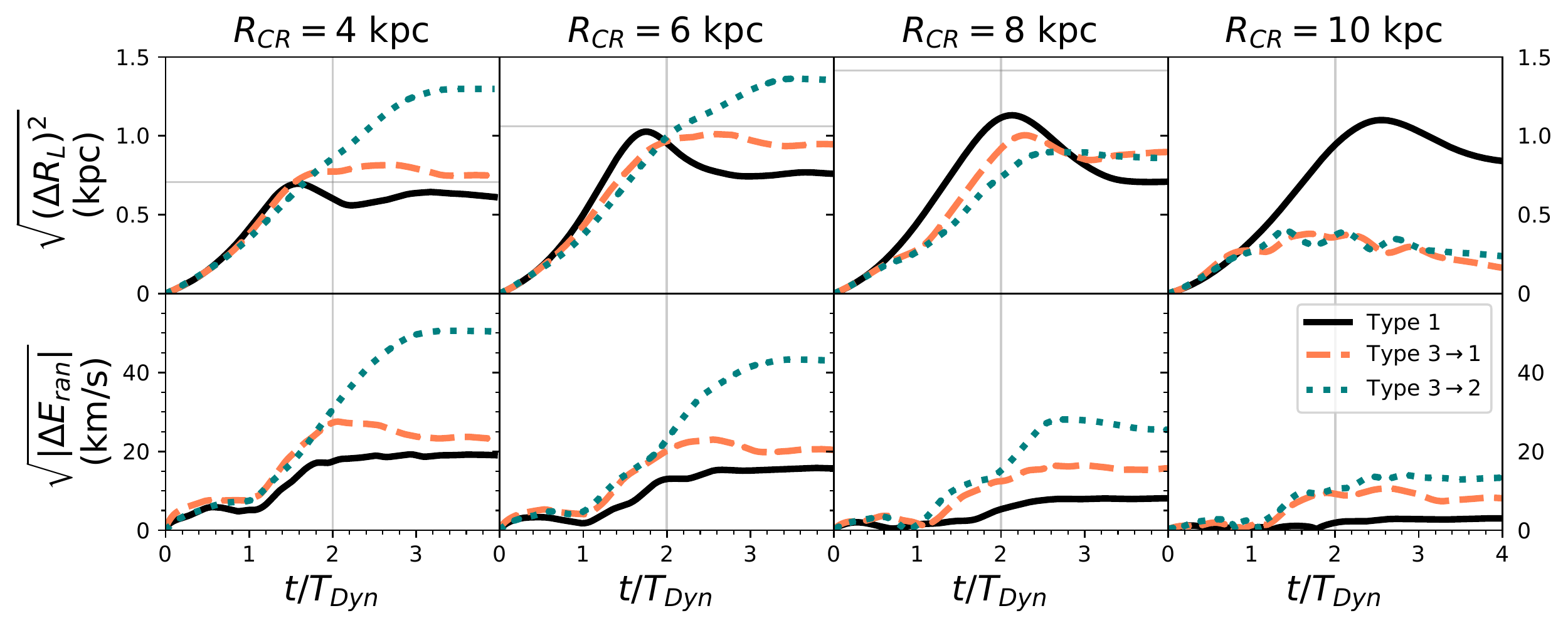}
    \caption{Time evolution of the orbital angular momentum and random orbital energy for populations of star particles in three classification categories.  These models use an exponential surface density profile ($\Sigma(R)\propto e^{-R}$) to determine the spiral amplitude.  Orbital classifications shown are Type~1 (always trapped -- solid, black), Type~3$\rightarrow$1 (resonant overlap occurs, but remain trapped -- dashed, orange), and Type~3$\rightarrow$2 (resonant overlap occurs and no longer trapped -- dotted, teal).  Top panels show RMS changes in orbital angular momentum using mean orbital radius, $R_L$ as a proxy, \RMSL, where the horizontal lines indicate half the distance between \ULR s. Bottom panels show kinematic heating expressed in velocity units by using \DelEran.  Time is expressed in units of orbital periods, $T_{\rm Dyn}$, where the vertical grey line indicates the moment of peak spiral amplitude.}
    \label{fig:5_expR}
\end{figure*}

\begin{figure*}
    \includegraphics[width=2\columnwidth]{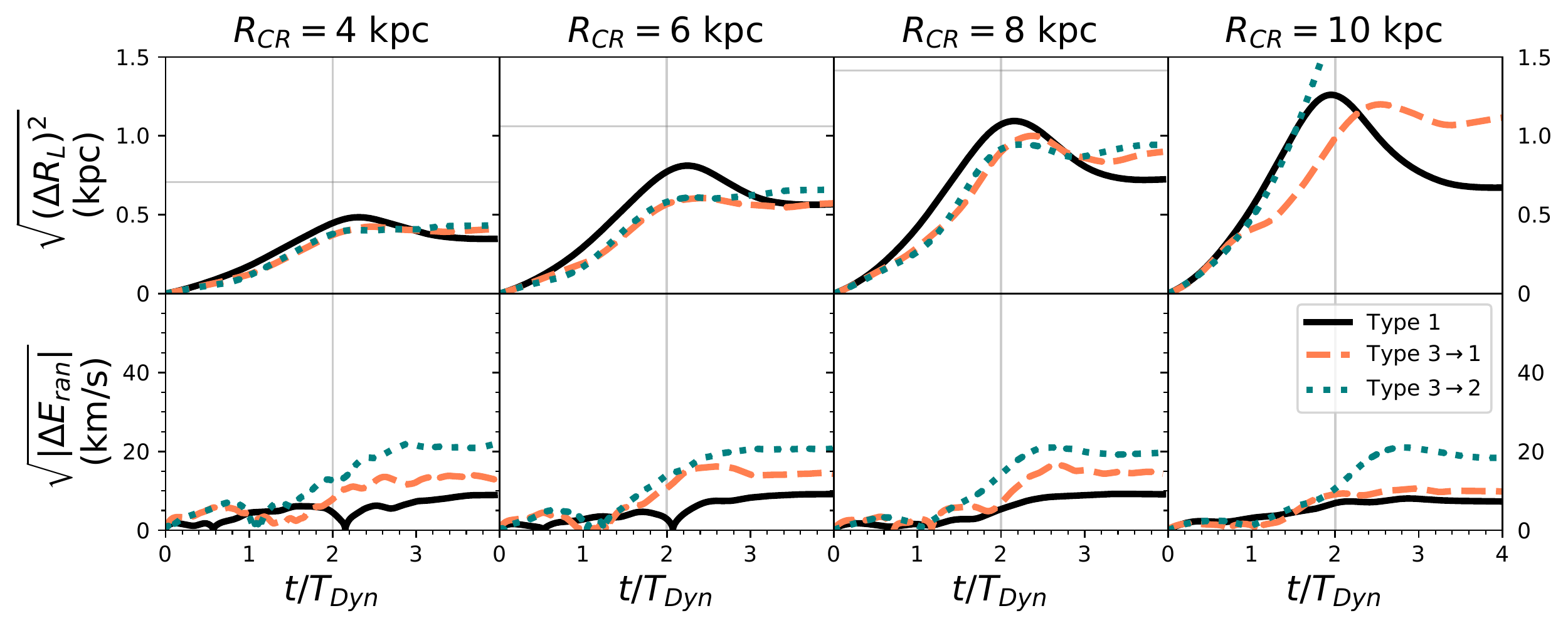}
    \caption{Plots and curves have the same meaning as in Figure~\ref{fig:5_expR}, where this figure is for models that use an inverse radial surface density profile ($\Sigma(R)\propto R^{-1}$) to determine the spiral amplitude.}
    \label{fig:5_invR}
\end{figure*}

Figures~\ref{fig:5_expR} and~\ref{fig:5_invR} show the time evolution for the changes in the orbital angular momentum and random orbital energy for populations of star particles in three classification categories.  Figure~\ref{fig:5_expR} shows models using an exponential surface density profile to determine the spiral amplitude and Figure~\ref{fig:5_invR} uses an inverse radial profile.  Orbital classifications shown are Type~1 (always trapped -- solid, black), Type~3$\rightarrow$1 (resonant overlap occurs, but remain trapped -- dashed, orange), and Type~3$\rightarrow$2 (resonant overlap occurs and no longer trapped -- dotted, teal).  Top panels show RMS changes in orbital angular momentum, \RMSL, using mean orbital radius, $R_L$, as a proxy.  Horizontal lines indicate half the distance between \ULR s as expressed in equation~\ref{eqn:DeltaULR}. Bottom panels quantify kinematic heating as the square root of the absolute value of the sum of the changes in random orbital energy, \DelEran, so that these changes are expressed in units of velocity.  Time is expressed in units of orbital periods, $T_{\rm Dyn}$, where the vertical grey line indicates the moment of peak spiral amplitude.

Trapped orbits (Type~1) have an increase in \RMSL\  over time and with increasing spiral amplitude in each model.  In most cases, the maximum value for \RMSL\ occurs nearly concurrently with the peak spiral amplitude (vertical, grey line).  One exception is the offset of this peak for Type~1 (always trapped) orbits, which likely reflects a 
time-scale difference between the imposed spiral lifetime governed by $\sigma_t$ (equation~\ref{eqn:et}) in these simulations and the time-scale for self-gravitational transient spiral structure. \Ct\,likely plays a role in the {self-regulation of transient spiral amplitude} \citep{SB02}.  Further, the time-scale for the radial oscillations of a trapped orbit depends on the maximum radial excursion for that orbit.  For a population of trapped stars, the peak in \RMSL\ likely occurs on the time-scale that maximizes the phase mixing of trapped orbits, where that time-scale changes with the artificially imposed spiral growth.  Relevant to this study is the amplitude of the peak in \RMSL, which is robust whether or not the spiral lifetime is artificially imposed.

The degree of kinematic heating for trapped orbits is not negligible, but is relatively insignificant when compared to other orbital classifications. The low grade heating of Type~1 orbits likely arises from a combination of factors including interactions with spiral arms away from \CR\,and near passes with the \ULR s without crossing.  Type~1 orbits experience the greatest degree of heating in models with \CR\,closer to the inner disk when the surface density profile is exponential. This same demographic describes models with a greater degree of resonant overlap and, therefore, correlates with the fraction of Type~1 orbits that have near encounters with an \ULR.  

Trapped orbits that experience resonant overlap (Type~3) also have a rise in \RMSL, but these changes are associated with kinematic heating.  In all cases, orbits that are not trapped after experiencing resonant overlap (Type~3$\rightarrow$2) have larger changes in both \RMSL\ and \DelEran\ than those orbits which remain in trapped orbits (Type~3$\rightarrow$1).  Nonetheless, \textit{trapped orbits that experience resonant overlap are \kmy\,heated} regardless of whether or not they remain in trapped orbits.  With the exception of M10e, \textit{Type~3 orbits have larger \RMSL\ by the end of the simulation than their Type~1 counterparts}.  The outstanding case of M10e is likely to do with the very small fraction of Type~3 orbits (2.3\%). 

\subsection{Constraints on radial excursions}\label{s:DeltaRmax}

The large fraction of orbits that are not trapped after an episode of resonant overlap (Type~3$\rightarrow$2/Type~3) suggests that this dynamical response could be important to understand how disks evolve around \CR. The distance between \ULR s increases with increasing radius of \CR\,and decreasing number of spiral arms (equation~\ref{eqn:DeltaULR}).  This study is limited to spirals with $m=4$ symmetry and so explores the consequence of the spacing between the \ULR s as a function of radius of \CR\,only.  

\Rm\,from a single episode of \ct\ depends on the radial range of the \CR\,region \citep{DW15}, which scales with the strength of the spiral perturbation \citep{SB02}.  The \CR\,region is radially more broad toward the galactic center for spirals of the same fractional amplitude of exponential surface density profile.  In models that use an inverse radial surface density profile, the radial range for the \CR\,region closely follows the trend in the radial distance between \ULR s.  The following discussion does not argue a preference for an underlying model.  Rather, it explores the role resonant overlap has on limiting the radial extent of cold \RM\ from \ct.  

Figures~\ref{fig:potentials_expR} and~\ref{fig:potentials_invR} illustrate how the difference between these two models for spiral amplitude affect the fraction of the \CR\,region that overlaps with the \ULR s.  
The horizontal lines in figures~\ref{fig:5_expR} and~\ref{fig:5_invR} show the distance between the \CR\,radius and an \ULR\,(solid, gray).
The curves for \RMSL\ for Type~1 orbits never exceed $\Delta R^{(2)}_{\rm LR}/2$ (equation~\ref{eqn:DeltaULR}) even when the radial range for the \CR\ region is greater than the annular range between the \ULR s.
There is therefore a limit to the RMS change in orbital angular momentum, and thus radial changes, for a population of stars migrating from \ct\ that is set by the spacing between \ULR s. The spacing between \ULR s is more restrictive toward the galactic center.  The linear approximation that maximum changes in orbital angular momentum from \ct\ are set by spiral strength must be further constrained by the non-linear response at resonant overlap.  

Trapped orbits that experience resonant overlap but remain trapped (Type~3$\rightarrow$1) are also limited by the constraint that \RMSL$\,\leq \Delta R^{(2)}_{\rm LR}/2$.  However, orbits that begin trapped at \CR\ but end in non-trapped orbits after an episode of resonant overlap (Type~3$\rightarrow$2) can have changes in orbital angular momentum that exceed this limit.  The implication is that strong transient perturbations can induce large changes in \RMSL, but \textit{these changes could be dominated by a large fraction of stars that are \kmy\,heated due to resonant overlap, even when their changes in orbital angular momentum are around \CR.}

It is worth considering that the trends shown in Figure~\ref{fig:4b} include regions at lower \gc\ radii where, in the Milky Way, the kinematics would be dominated by the bar. Assuming a circular velocity of 220~km~s$^{-1}$ and bar pattern speed $41\pm 3$~km~s$^{-1}$~kpc$^{-1}$ \citep{Bovy19}, the Milky Way's bar has \CR\ radius equal to $\sim 5.4$~kpc \cite[see also][]{WGP15,PGW17}.  The shape of the underlying potential, and resulting shape of the \CR\ region, for a bar is rather different from the case of a spiral pattern.  Additionally, the vertical component cannot be taken to be separable as can be done with a spiral in a thin disk and as is assumed in the formulation of the capture criterion \citep{DW15}.  Nonetheless, there is a significant degree of resonant overlap between the \ULR s and the \CR\ region for a bar.  The regions around the L4 and L5 points, which are at the \CR\ radius and coincident with the azimuthal line through the bar's minor axis, are typically considered stable.  However, in a given barred disk there may be additional resonances between the galactic center and the \CR\ radius of the bar. In this scenario orbits that would be considered stable in the approximation that \CR\ is at a particular radius would be subject to resonant overlap when recognizing that the \CR\ resonance fills some volume in phase-space.  Such resonant overlap would presumably induce a chaotic response, where chaotic regions in phase-space are expected to be depopulated \citep{Pfenniger90}.  Indeed, \cite{Buta17} uses the evacuation of orbits in these regions in the so called \lq gap method' to infer the location of \CR\ and thus other resonances of the bar.  These dynamics are currently under further investigation.  This discussion is relevant in the current study only insofar as to recognize that the kinematic trend does not reflect the kinematics from resonant overlap from a bar.

\section{Conclusions}\label{s:conclusion}

\Ct\ is a resonant effect that happens around the radius of \CR\ with a transient spiral pattern.  
A critical underlying assumption for \RM\ by \ct\ is that each transient spiral rearranges orbital angular momentum around \CR\ without causing kinematic heating.
Multiple generations of transient spirals with a range of pattern speeds could cause stellar mean orbital radii to radially diffuse from their birth place in a random walk-like fashion. The standard deviation for the final radial distribution of migrated stars is proportional to the size of each step.
To first approximation, the radial size of each step increases with spiral strength \citep{SB02}.  
This study aims to constrain the assumption that a single episode of \RM\ from \ct\ can happen across arbitrarily large distances (equation~\ref{eqn:Rmax}) and remain \kmy\ cold.  A clear upper limit on the step size is a constraint on the rate of random walk-like diffusion of stars across the disk on a given timescale from \ct.
%since the \CR\ region is bounded by \ULR s.  

A suite of models is populated with initial conditions that are generated from a distribution function ($f_{\rm new}$, equation~\ref{eqn:fnew}) designed to produce a flat rotation curve and an exponential surface density profile in a \kmy\ warm 2D disk.  Each set of initial conditions is integrated over four orbital periods (4$T_{\rm Dyn}$) through a spiral disk potential with underlying logarithmic potential selected to reproduce a flat rotation curve and a superposed density wave spiral pattern.  Each spiral pattern has a \CR\ radius set to be between $4-10$~kpc.  Spirals have a Gaussian time-dependent amplitude with peak amplitude at $t=2 T_{\rm Dyn}$ and lifetime set by standard deviation $\sigma_t=T_{\rm Dyn}$.  Peak spiral amplitudes have radial dependence based on surface density profiles that follow either an exponential or inverse radial form.

A single spiral pattern could produce trapped orbits around the \CR\ resonance that are also resonant with the \ULR s from the same spiral pattern \citep{DW15}.  
Each resonance is governed by different defining frequencies thus inducing a chaotic dynamical response \citep{SD19}.
Populations of trapped orbits that experience resonant overlap, compared to trapped orbits that do not, have larger changes in their orbital angular momentum and are  \kmy\ heated.  Approximately $60\pm20\%$ of the orbits subject to resonant overlap change in orbital type from trapped to non-resonant with \CR.  Model results suggest that, when resonant overlap is possible, the largest changes in orbital angular momentum for stars subject to \ct\ are also significantly \kmy\ heated.
This contradicts the assumption that \RM\ through \ct\ is necessarily a cold process.  

The distance between the \ULR s sets an upper limit on the radial range for \RMSL\ from cold \RM\ from \ct.  
Resonant overlap induces kinematic heating in trapped stars that have mean orbital radius ($R_L$) cross an \ULR\ setting an upper limit to
\RMSL$\,< R^{(2)}_{\rm LR}/2$ for \kmy\ cold \ct.  In cases when the expected maximum radial excursions are greater than the distance between the \ULR s ($\Delta R_{\rm max}>R^{(2)}_{\rm LR}$) \RM\ is a \kmy\ heating process around \CR.
It is not expected that resonant overlap would be important for weak spiral patterns, since the radial range of the \CR\ region is correlated with spiral strength.

The case study of an exponential disk with a flat rotation curve is used to illustrate scaling relations that can be drawn between the radius of \CR\ and the role of resonant overlap.  In this case, the distance between the \ULR s decreases linearly with decreasing radius of \CR\ while the radial range of the \CR\ region increases.  Thus, for the same fractional amplitude for the spiral strength, the boundary for resonant overlap is more restrictive toward the galactic center causing a large degree of kinematic heating at small \gc\ radii.  
The flip side of this is that \ct\ is expected to remain \kmy\ cold toward the outer disk.

Future work is in progress to explore the role of resonant overlap on kinematic heating near \CR\ from multiple generations of spiral patterns.  A 3D simulation is necessary in order to investigate how strong radial mixing with moderate, correlated kinematic heating from \ct\ affects vertical disk kinematics and thickening. 

\section*{Acknowledgements}

The authors thank the anonymous referee for comments that helped clarify the manuscript.
KD would like to thank Ortwin Gerhard for useful conversations during the course of this study.  This work was, in part, performed at the Aspen Center for Physics, which is supported by National Science Foundation grant PHY-1607611.  
DS is funded by the DOE through the NSF-DOE Partnership in Basic Plasma Science and Engineering grant DE-SC0018258 and by an NSF Early Career Award PHY-1846943.
%\kd{Do the software acknowledgement thing}



\bibliographystyle{aasjournal}
\bibliography{Main} % if your bibtex file is called example.bib

%%%%%%%%%%%%%%%%%%%%%%%%%%%%%%%%%%%%%%%%%%%%%%%%%%

\end{document}